\documentclass{article}

\usepackage{arxiv}

\usepackage[utf8]{inputenc} 
\usepackage[T1]{fontenc} 
\usepackage{hyperref} 
\usepackage{url} 
\usepackage{booktabs} 
\usepackage{amsfonts} 
\usepackage{nicefrac} 
\usepackage{microtype} 
\usepackage{cleveref} 
\usepackage{lipsum} 
\usepackage{graphicx}
\usepackage{doi}

\usepackage{floatrow}
\usepackage{subfig}
\usepackage{tabularx}

\usepackage{tikz}
\def\checkmark{\tikz\fill[scale=0.4](0,.35) -- (.25,0) -- (1,.7) -- (.25,.15) -- cycle;} 

\title{Artificial neural networks for magnetoencephalography: a review of an emerging field}

\author{
 Arthur Dehgan \\
 MILA, CoCo Lab\\
 Université de Montreal\\
 \texttt{dehganar@mila.quebec}\\
 \textit{(Corresponding author)}\\
 \And
 Hamza Abdelhedi\\
 MILA, CoCo Lab\\
 Université de Montreal\\
 \texttt{hamza.abdelhedi@umontreal.ca} \\
 \And
 Vanessa Hadid\\
 McGill University Health Centre\\
 McGill University\\
 \texttt{vanessa.hadid@mcgill.ca} \\
 \And
 Irina Rish \\
 MILA\\
 Université de Montreal\\
 \texttt{rish@iro.umontreal.ca} \\
 \And
 Karim Jerbi \\
 MILA, CoCo Lab\\
 Université de Montreal\\
 \texttt{karim.jerbi@umontreal.ca} \\
}

\begin{document}
\maketitle

\begin{abstract}
\textit{Objective:} Magnetoencephalography (MEG) is a cutting-edge neuroimaging technique that measures the intricate brain dynamics underlying cognitive processes with an unparalleled combination of high temporal and spatial precision. While MEG data analytics have traditionally relied on advanced signal processing and mathematical and statistical tools, the recent surge in artificial intelligence (AI) has led to the growing use of machine learning (ML) methods for MEG data classification. An emerging trend in this field is the use of artificial neural networks (ANNs) to address various MEG-related tasks. This review aims to provide a comprehensive overview of the state of the art in this area. \textit{Approach:} This topical review included studies that applied ANNs to MEG data. Studies were sourced from PubMed, Google Scholar, arXiv, and bioRxiv using targeted search queries. The included studies were categorized into three groups: 'Classification', 'Modeling', and 'Other'. Key findings and trends were summarized to provide a comprehensive assessment of the field.\textit{Main results:} We identified 119 relevant studies, with 70 focused on 'Classification', 16 on 'Modeling', and 33 in the 'Other' category. 'Classification' studies addressed tasks such as brain decoding, clinical diagnostics, and BCI implementations, often achieving high predictive accuracy. 'Modeling' studies explored the alignment between ANN activations and brain processes, offering insights into the neural representations captured by these networks. The 'Other' category demonstrated innovative uses of ANNs for artifact correction, preprocessing, and neural source localization.
\textit{Significance:} By establishing a detailed portrait of the current state of the field, this review highlights the strengths and current limitations of ANNs in MEG research. It also provides practical recommendations for future work, offering a helpful reference for seasoned researchers and newcomers interested in using ANNs to explore the complex dynamics of the human brain with MEG.
\end{abstract}


\newpage
\section{Introduction}

Artificial neural networks (ANNs) have had an enormous impact on most research fields, especially in computer vision and natural language processing (\cite{dlbook}). In recent years, ANNs have become increasingly used for brain data analyses and modeling, yielding ample evidence for their added value in neuroscience research (\cite{richards2019deep, zador2022toward}). Deep learning (DL) has been applied in the analysis of various types of brain imaging modalities, including structural magnetic iesonance imaging (MRI), functional magnetic resonance imaging (fMRI), electroencephalography (EEG), and magnetoencephalography (MEG). While using ANNs specifically with EEG has been the subject of comprehensive reviews (e.g. \cite{walther2023systematic, roy2019deep}), using ANNs with MEG data is less common, and the current state-of-the-art has yet to be surveyed. This combination, however, is of growing interest. MEG provides direct, time-resolved recordings of brain activity, which can be used to compare the dynamics and internal representations of ANN models to those observed in the brain. For ANN researchers, this opens up possibilities for testing model outputs against high-temporal-resolution biological signals and for training models that simulate aspects of cognitive processing. For MEG researchers, ANNs offer powerful tools to classify different brain states, uncover hidden patterns in brain activity, and model the complex structure of MEG datasets. These shared opportunities have led to an increasing number of studies exploring how MEG and ANNs can be integrated in practice. The present review seeks to fill in this gap by providing a detailed survey of the main applications of ANNs in MEG research to date. The strengths and limitations of the approaches reviewed allow us to highlight the potential of ANNs in future MEG research and the main challenges in this field.

\subsection{Magnetoencephalography and brain imaging}

MEG is a non-invasive electrophysiological brain imaging technique that detects magnetic signals generated by synchronized neuronal ensembles in the brain, typically recorded with approximately 100--300 sensors depending on the system. Although it shares similarities with EEG, such as electrophysiological signal origins and millisecond-range temporal resolution, MEG offers distinct advantages over EEG and other modalities like fMRI. Unlike EEG, MEG is less susceptible to spatial smearing from variations in conductivity across brain tissues and the skull, due to its sensitivity to magnetic permittivity changes, which are relatively homogeneous in the brain (\cite{singh2014magnetoencephalography}). This makes MEG particularly well-suited for source reconstruction and analyzing spatio-temporal brain dynamics in cortical source space (\cite{gavaret2014meg, baillet2001electromagnetic}). In contrast to fMRI, which measures hemodynamic responses only indirectly reflecting neural activity with a temporal resolution of several seconds, MEG directly captures electrophysiological activity on a millisecond scale (\cite{hall2014relationship}).

Overall, with its high temporal resolution and improved spatial precision over EEG, MEG excels in studying complex brain dynamics, despite challenges like field spread that source reconstruction mitigates but does not fully resolve (\cite{kaur2022recent}). Both EEG and MEG produce sensor-level data that are highly intercorrelated and have limited spatial interpretability. These ambiguities can be reduced with MEG by applying source localization techniques (\cite{baillet2001electromagnetic, kaur2022recent}). Compared to fMRI, another widely used functional neuroimaging technique, MEG’s millisecond-scale resolution provides a significant advantage for capturing rapid neural events, whereas fMRI’s lower sampling rate limits its temporal precision (\cite{hall2014relationship}). Although source reconstruction does not entirely resolve the field spread problem or linear mixing issue, it facilitates exploration of the functional role of specific brain areas with more anatomical precision than sensor-level analysis (\cite{da2013eeg}).

Given its high temporal and spatial resolution, MEG has become an established imaging modality particularly suited for addressing neuroscientific inquiries involving intricate spatial, temporal, and spectral patterns of distributed brain dynamics. These include the real-time integration of information from one or multiple modalities and the neural dynamics underpinning advanced cognitive functions such as language processing and decision-making. For a complete overview of the advantages and drawbacks of MEG in neuroscience, see \cite{baillet2017magnetoencephalography}.

\subsection{Standard M/EEG analysis pipeline}
Generally, the M/EEG analysis pipeline consists of a sequence of processing steps that start from the sensor-level raw data and, after a series of manipulations, ultimately yield observations that address a research question. Standard preprocessing pipelines often include data cleaning and artifact rejection procedures, including channel rejection, notch-filtering, and independent component analysis (ICA \cite{hyvarinen2000independent}).

M/EEG recordings yield rich and high-dimensional data, and typically, the analysis pipeline involves computing a set of variables from the raw data, also known as hand-crafted feature extraction in the case of ML analysis. Computing these variables (or features) relies on processing pipelines that can be broadly divided into two main categories: time-domain or frequency-domain analyses (which can also be jointly explored using time-frequency representations of the data (\cite{cohen2014analyzing, buzsaki2006rhythms}).

Time domain analyses typically involve the computation of features such as event-related potentials (ERPs \cite{berger1929elektroencephalogramm}), long-range temporal correlations (LRTC) using detrended fluctuation analysis (DFA \cite{peng1995long} ), complexity measures (e.g. Hjorth parameters \cite{hjorth1970eeg}, Hurst exponent \cite{hurst1951long}, fractality measures (fractal dimension) or entropy measures (approximate entropy \cite{pincus1991approximate}, sample entropy \cite{richman2000physiological}, permutation entropy \cite{bandt2002permutation}). Frequency-domain analyses typically employ spectral analysis methods, such as the wavelet (\cite{mallat1989theory}) or Fourier transform (\cite{welch1967use, cohen2014analyzing}), to derive features like the spectral power across canonical frequency bands (e.g. delta, theta, alpha, beta, gamma). 

In time- or frequency-domain analyses, connectivity features can be calculated using multi-channel or whole-brain source data. Typical connectivity metrics assess different associations between signals from various sources or sensors. These metrics include magnitude-squared coherence (MSCoh \cite{fries2005mechanism}), phase locking value (PLV \cite{lachaux1999measuring}), phase-lag index (PLI \cite{stam2007phase}), and weighted PLI (wPLI \cite{vinck2011improved}), to name but a few.
Different features capture complementary -although sometimes partly overlapping- aspects of the neural dynamics associated with various behavioral states.

Sensor-level analyses in neuroscience are informative, but source reconstruction is essential for improving the interpretation of the spatial distribution of significant effects. This step involves defining a forward model and solving the inverse problem ( \cite{hamalainen1993magnetoencephalography} ). Effective source reconstruction allows for precise identification of neural markers in specific brain regions, enhancing our understanding of complex brain functions and informing both hypothesis validation and exploratory data-driven research.

\subsection{Machine learning versus inferential statistics} 

A typical M/EEG neuroscience study employs either inferential statistics, using hypothesis testing methods such as t-tests or ANOVA, or statistical learning approaches, notably machine learning (ML). The choice between these approaches depends on the specific research questions, with inferential statistics often used to test predefined hypotheses about neural activity, while ML is typically applied to explore complex patterns and predictions from high-dimensional datasets. Both approaches have significantly contributed to progress in neuroscience, as detailed in \cite{bzdok2017classical}. Additionally, the bayesian framework has emerged as a complementary avenue in M/EEG analysis (\cite{wipf2009unified}), allowing hypothesis testing from a probabilistic perspective. While hypothesis testing, whether conducted through traditional methods or bayesian approaches, has long been a staple in neuroscience, it remains the subject of ongoing debate and scrutiny (\cite{szucs2017null}). Despite its undeniable strength and utility, null hypothesis significance testing (NHST) has several limitations. First, it heavily relies on a priori choices guided by literature and expert domain knowledge. Additionally, it assumes generalization of findings based solely on the representativeness of the studied sample to the target population. As an alternative approach to examining hand-picked variables informed by the literature, MEG researchers have begun to use a variety of data-driven approaches. These broadly include two types of approaches to relevant feature identification with hardly any a priori information: (a) Feeding a large number of hand-crafted features into a decoding framework and assessing their relative contributions to model performance, and (b) Learning the relevant features from the data themselves, a form of data-driven automatic feature extraction known as representation learning. Unlike simpler algorithms such as support vector machines (SVM), linear discriminant analysis (LDA), or decision trees which generally rely on hand-crafted features, ANNs can automatically learn and extract relevant features from data. Despite its conceptual appeal, representation learning comes with its limitations, in particular when it comes to interpreting the learned features. ANNs are often criticized for being 'black boxes' compared to more classical ML algorithms, which use hand-crafted features as inputs. Enhancing the feature interpretability is an active research topic in deep representation learning.

\subsection{Artificial neural networks in a nutshell}

Understanding ANNs begins with grasping the concept of linear regression, where the output \(y\) is given by \(y = wx + b\), with \(w\) representing the weight, \(x\) the input, and \(b\) the bias. In ANNs, this equation gets extended by an activation function \(f\), transforming it to \(y = f(wx + b)\) (\cite{seber2003linear, rosenblatt1958perceptron}). As the neural network gains depth through additional layers, these equations combine to form increasingly complex mathematical models. A neural network's complexity, or \textit{capacity} (\cite{hastie2009elements}), indicates its ability to model intricate functions, but this can be a double-edged sword. High capacity often leads to \textit{overfitting} (\cite{hastie2009elements}), where the model learns the training data too well, capturing noise and inaccuracies, thus performing poorly on new data. To mitigate overfitting, regularization techniques such as \textit{maxpooling} (\cite{krizhevsky2012imagenet}), which reduces feature map dimensions, \textit{dropout} (\cite{srivastava2014dropout}), which nullifies a subset of neurons during training, and \textit{batch normalization} (\cite{ioffe2015batch}), which standardizes layer outputs, can be applied. Another regularization method comes in the form of \textit{data augmentation} (\cite{shorten2019survey}), where the initial dataset can be enlarged by introducing noise to the data, rotating, or cropping it. Learning in ANNs occurs through optimization, typically employing stochastic gradient descent (SGD \cite{robbins1951stochastic}). During each training iteration, the model's parameters are adjusted based on computed gradients from a loss function, facilitated by a technique known as \textit{backpropagation} (\cite{rumelhart1986learning}). As we progress to more advanced ANNs, we encounter \textit{convolutional neural networks} (\cite{lecun1998gradient}) optimized for image data, \textit{recurrent neural networks} (\cite{hochreiter1997long}) ideal for sequences, and \textit{attention models} (\cite{vaswani2017attention}) that allocate varying degrees of focus on different parts of the input. Another important class includes autoencoders, which are typically unsupervised networks designed to learn efficient data codings. They consist of an encoder compressing the input to a latent representation and a decoder reconstructing the input from this representation, aiming to capture essential data features. Autoencoders can utilize various architectures like CNNs or RNNs. Sparse autoencoders (SAEs) are a variant that adds sparsity constraints to encourage compact representations. Overall, the domain of ANNs extends from fundamental mathematical principles to intricate architectures and sophisticated methods for effective training.

While the ANNs described in this review originate from early inspiration in neuroscience, it is important to distinguish them from biologically detailed models used in computational neuroscience. Models such as integrate-and-fire neurons, Hebbian learning mechanisms, spiking neural networks, and neural mass models aim to capture the physiological behavior of neural systems with varying levels of abstraction. In contrast, the ANNs considered in this review are primarily optimized for predictive performance and scalability in ML applications. This distinction highlights the divergence between biologically motivated modeling and task-driven computational approaches — both valuable, but grounded in different research priorities (\cite{cohen2022recent}).


\subsection{ANNs for neuroimaging data}

Recently, ML has garnered much interest from the neuroscience research community (\cite{mlneurosystems, mlneurodecoding, mlneuro}). While deep learning has predominantly been applied to fMRI and MRI data due to their compatibility with existing architectures, its use in EEG analysis is also growing, partly due to EEG's wider availability. However, the unique strengths of MEG, such as its high temporal and spatial resolution, present a compelling case for its integration with advanced techniques like ANNs. These networks are well-suited to handle MEG's complex, high-dimensional data. They apply to various tasks, including classification, data modeling, representation learning, data cleaning, and source estimation, thereby contributing to a more nuanced understanding of brain functions. The use of ANNs to analyze MEG data is gaining traction, offering novel perspectives and data analyses that standard MEG methodologies typically do not provide. Although arguably still in the early stages, exciting progress has already been reported in this emerging field.

\subsection{Goal of this review}

 A growing number of AI researchers are turning to MEG as a rich source of non-invasive brain-wide neural data, recognizing its high spatio-temporal resolution and ability to capture complex cognitive dynamics as key assets for advancing more efficient, robust, and brain-inspired neural network models. This review provides a comprehensive survey of ANN applications in MEG data analysis, highlighting their benefits and potential added value. These techniques have proven to be shown to be useful in various neuroscience contexts. By providing examples of architectures that emulate aspects of visual processing and techniques, we aim to provide an up-to-date overview of the demonstrated the potential of this approach in ANN-based MEG studies. We hope to expedite researchers' path to developing their own implementations. More globally, we hope this review will encourage researchers well-versed with MEG to integrate deep learning techniques into their work, and conversely encourage experts in deep learning with an interest in neuroscience to consider MEG as a particularly promising brain measurement modality in the context of Neuro-AI research applications.

\section{Methods}
\label{sec:methods}

\subsection{Litterature research}
\begin{figure}[htbp]
 \centering
 \includegraphics[width=.9\linewidth]{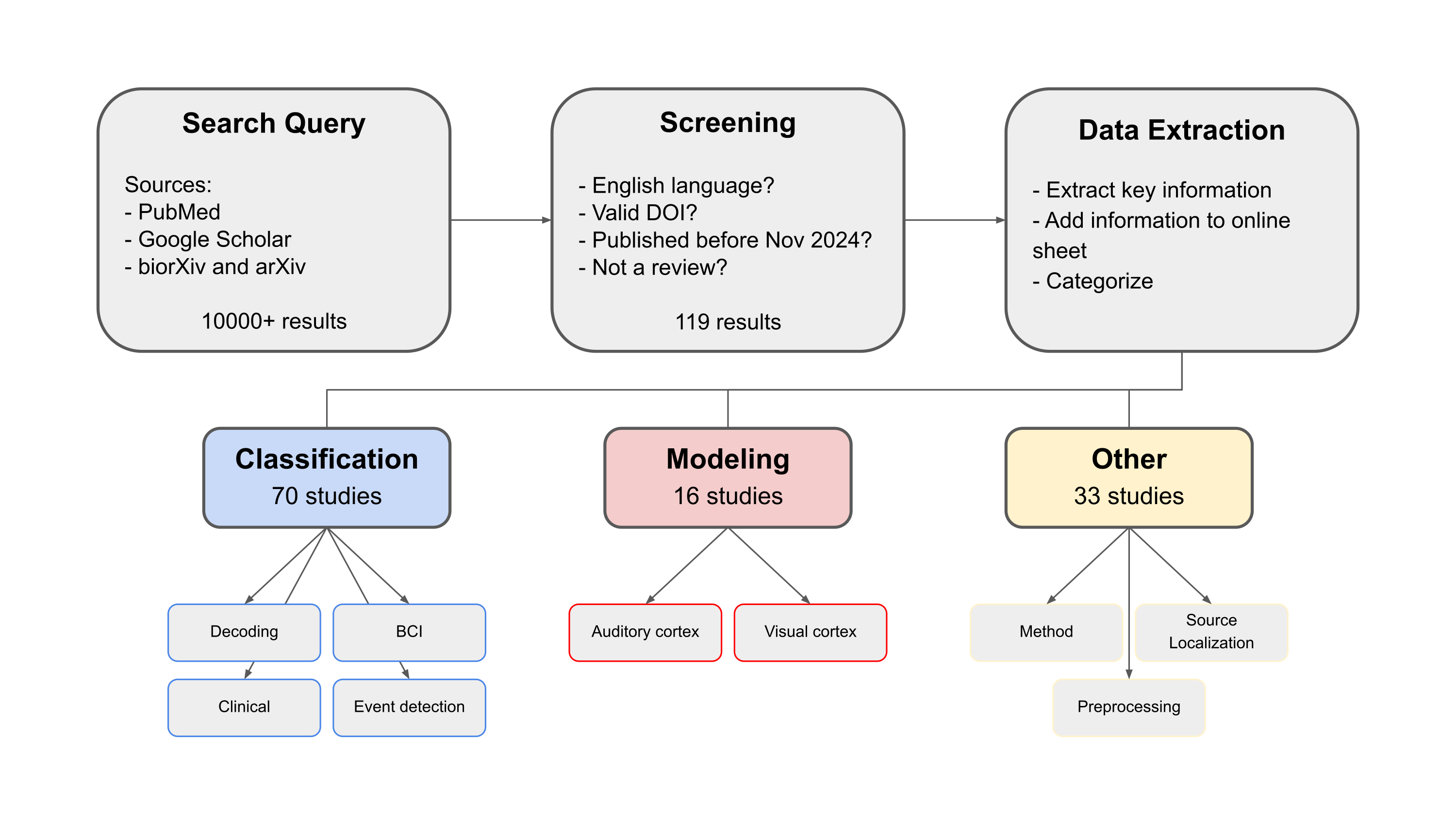}
 \caption{Schematic overview of the literature review process. An initial search query across specified databases yielded over 10,000 results. These were screened based on language, DOI availability, publication date (< Nov 2024), and exclusion of reviews, resulting in 119 included studies. Following data extraction, studies were categorized into three main groups: 'Classification' (N=70; further subdivided into decoding, BCI, clinical, and event detection), 'Modeling' (N=16; subdivided by cortical focus), and 'Other' methodological applications (N=33; subdivided into method, source localization, and preprocessing).}
 \label{fig:review}
\end{figure}

For this review, we collected English-language conference and journal papers using PubMed, Google Scholar, arXiv, and bioRxiv. We used the following query to find the papers referenced in this review: ("deep learning" OR "artificial neural network*" OR "Convolutional neural net*" OR CNN OR "Recurrent neural net*" OR RNN) AND ("MEG" OR Magnetoencephalogra*). Relevant papers were selected by examining the title first, followed by the abstract. Finally, we used a PDF search for the terms from the query to better understand how they are used. 
This study includes conference papers, journal papers, and pre-prints. The final inclusion criterion requires papers to be primary research articles (i.e., not reviews), possess a valid digital object identifier (DOI) and a publication date preceding November 2024. Figure \ref{fig:review} provides a schematic overview of the literature search and selection process employed in this review.

Tables \ref{tab:classification}, \ref{tab:classification2}, \ref{tab:modeling}, \ref{tab:other} provide an overview of the key information extracted from the included studies for this survey. More detailed information on each paper was collected in a table we have made available online as a Google sheet: \url{https://tinyurl.com/ub3s5mr}. See table \ref{tab:description} for details on each information category extracted from the papers.

\begin{table}
 \centering
 \begin{tabularx}{\textwidth}{l|X|l|l}
 Item & Description & Online table & Included table\\
 \hline\hline
 Author & Name of the main author(s). & \checkmark & \checkmark \\ 
 Publication year & The year the paper/pre-print was published.& \checkmark & \checkmark \\
 Title & The title of the paper.& \checkmark & \\
 DOI & The DOI of the paper.& \checkmark & \\
 Category & The paper category (classification, modeling, or other).& \checkmark & \checkmark \\
 Subcategory & Subcategory of paper.& \checkmark & \checkmark \\
 Goal & The goal of the study.& \checkmark & \\
 Subjects (N) & The number of subjects for the study.& \checkmark & \checkmark \\
 Notes on subjects & Information and details on subjects & \checkmark & \\
 Sample size (trials) & The total number of samples used.& \checkmark & \checkmark \\
 Notes on sample size & Information and details on the number of samples. & \checkmark & \\
 Input data & The type of input data given to the neural network.& \checkmark & \\
 Detailed input data & A more detailed version of the input data column.& \checkmark & \\
 Data augmentation & Data augmentation techniques used, if any. & \checkmark & \\
 Validation & The type of validation or model selection technique used.& \checkmark & \checkmark \\
 Sensors & The number of sensors used.& \checkmark & \\
 SF	& The sampling frequency used.& \checkmark & \\
 Performance & The performances that the paper was able to reach.& \checkmark & \\
 Trial length & Length of a single data point given to the neural network.& \checkmark & \\
 Baseline & Previously attained/baseline performances.& \checkmark & \\
 Architecture (depth) & The type of architecture used and its depth.& \checkmark & \checkmark \\
 Detailed architecture & The detailed architecture of the network used.& \checkmark & \\
 preprocessing & The detailed preprocessing steps used.& \checkmark & \\
 Processing & Data processing/feature extraction process.& \checkmark & \\
 Specificities & Further details on how the study was led.& \checkmark & \\
 Interpretation & visualization techniques or interpretation tools, if any.& \checkmark & \\
 Training params & Details on the neural network training, when available.& \checkmark & \\
 Code availability & Whether the code is publicly available or not.& \checkmark & \checkmark \\
 \end{tabularx}
 \caption{Description of table items, and in which table they can be found. Full table and information can be found in \href{https://tinyurl.com/ub3s5mr}{https://tinyurl.com/ub3s5mr}}
 \label{tab:description}
\end{table}

\newgeometry{textwidth=7in}
\begin{table}
 \centering
 \resizebox{.95\textwidth}{!}{
 \newcolumntype{Y}{>{\raggedright\arraybackslash}X} 
\begin{tabularx}{\textwidth}{l|X|l|r|r|X|X|r@{}}
Year & Authors & Subcategory & N & trials & Architecture (depth) & Validation & Code\\
\hline
2016 & Yu et al. \cite{yu2016encoding} & Decoding& 15 & 2000& CNN (3) & train/test split & \\
2017 & Wang et al. \cite{wang2017towards} & Decoding& 2& 410 & MLP (1) & \begin{tabular}[t]{@{}l@{}}4Fold on train +\\holdout test set\end{tabular}& \\
2018 & Hramov et al. \cite{hramov2018artificial}& Decoding& 5& 150000& MLP (1) & train/valid split & \\
2018 & \begin{tabular}[t]{@{}l@{}}Frolov\\and Pisarchik \cite{frolov2018diagnostics}\end{tabular}& Decoding& 5& 10000 & MLP (1) & train/test split& \\
2019 & Dash et al\cite{dash2019decoding} & Decoding& 4& N/M & \begin{tabular}[c]{@{}l@{}}MLP (3)\\LSTM \end{tabular}& train/valid/test split & \\
2019 & Garry et al. \cite{garry2019classification}& Decoding& 605& 75396 & CNN (3) & train/valid/test split & \checkmark \\
2019 & Kim et al. \cite{kim2019canet}& Decoding& 27 & 5400& CNN + Attention (6) & 5Fold& \\
2019 & Kostas et al. \cite{kostas2019machine}& Decoding& 101& 3900& CNN and RCNN (N/A)& 5Fold & \\
2019 & Dash et al. \cite{dash2019towards}& Decoding& 8& 2400& MLP (1) & train/test split& \\
2019 & Huang and Yu \cite{huang2019cross}& Decoding& 16 & 588 & CNN (4) & train/test split& \checkmark \\
2020 & Abdellaoui et al. \cite{abdellaoui2020deep}& Decoding& 18 & N/M & CNN + Attention + LSTM (10) &train/test split & \checkmark \\
2021 & Dash et al. \cite{dash2021role}& Decoding& 7& 300 & MLP (1) & 5Fold& \\
2020 & Dash et al. \cite{dash2020neural}& Decoding& 10 & 180 & MLP (1) & 5Fold& \\
2021 & Li et al. \cite{li2021inter}& Decoding& 16 & 9414& 2 x GRU + FC& LOO & \\
2021 & Feng et al. \cite{feng2021new}& Decoding& 200& 397 & GoogLeNet - CNN (144) & train/valid split & \\
2021 & Chang et al. \cite{chang2021decoding}& Decoding& 8& 108 & GAN-like (5)& Holdout set & \\
2021 & Pilyugina et al. \cite{pilyugina2021comparing}& Decoding& 17 & 17& CNN (1) & train/test split & \\
2022 & Engemann et al. \cite{engemann2022reusable}& Decoding& 646& N/M & ShallowFBCSPNet + Deep4Net& 5Fold& \checkmark \\
2021 & Shi et al. \cite{shi2021categorizing}& Decoding& 17 & 10880 & EEGNet (4)& nested LOSO & \\
2023 & Zhang et al. \cite{zhang2023decoding}& Decoding& 20 & N/M &N/M& Leave One Out & \checkmark \\
2023 & Csaky et al. \cite{csaky2023group}& Decoding& 15 & 53100 & Wavenet Classifier (9)& LOSO& \checkmark \\
2023 & Özer et al. \cite{ozer2023brain}& Decoding& 18 & 9414& LSTM, GRU and CNN & 10Fold & \\
2023 & Bu et al. \cite{bu2023magnetoencephalogram}& Decoding& 12 & 8640& CNN (4) & 5Fold + train/test split& \checkmark\\
2024 & Boyko et al. \cite{boyko2024megformer}& Decoding& 96 & 2304000 &N/M& train/valid/test split& \checkmark \\
2024 & Yang et al. \cite{yang2024mad}& Decoding& 27 & 179977& N/A & train/valid/test split& \checkmark \\
2024 & Zubarev et al. \cite{zubarev2024robust}& Decoding& 12 & 3600& LF-CNN (3)& 9Fold& \checkmark\\
2024 & Yang et al. \cite{yang2024neugpt}& Decoding& 27 & 179977& NeuGPT & train/valid/test& \checkmark\\
2024 & Jayalath et al. \cite{jayalath2024brain}& Decoding& 900 & 2296800& CNN + LSTM & train/valid/test & \\
2020 & Shu and Fyshe \cite{shu2013sparse}& Decoding& 9 & 60 & SAE& LOO& \\
\end{tabularx}}
 \caption{Part 1 of 4 of the table containing condensed information about the included papers. This part includes all papers categorized as 'Classification' papers and subcategorized as 'Decoding' papers. More info can be found in the complete table: \href{https://tinyurl.com/ub3s5mr}{https://tinyurl.com/ub3s5mr}. N is the number of subjects included in the study. 'N/M' is used when specific information was either not mentioned or when it was impossible to infer from other information. N/A means the information does not apply to the study.}
 \label{tab:classification}
\end{table}
\restoregeometry

\newgeometry{textwidth=7.8in}
\begin{table}
 \centering
 \resizebox{.95\textwidth}{!}{
 \newcolumntype{Y}{>{\raggedright\arraybackslash}X}
\begin{tabularx}{\textwidth}{l|X|l|r|r|X|X|r@{}}
Year & Authors & Subcategory & N & trials & Architecture (depth) & Validation & Code \\
\hline
2018 & Meng et al. \cite{meng2018brain}& Clinical& 45 & N/M & CNN (4)& 4Fold & \\
2019 & Aoe et al. \cite{aoe2019automatic}& Clinical& 233& 70000 & CNN (12) & 10Fold& \checkmark \\
2020 & Gu et al. \cite{gu2020multi}& Clinical& 32 & N/M & MLP (4) + Self-attention head& cross-validation & \\
2020 & Zhang et al. \cite{zhang2020predicting}& Clinical& 32 & N/M & LSTM (2) & 10Fold + holdout test set & \\
2021 & Xu et al. \cite{xu2021graph}& Clinical& 129& N/M & G2G network (1) & train/valid split & \\
2021 & Giovannetti et al. \cite{giovannetti2021deep}& Clinical& 87 & 6525& AlexNet - CNN (5)& LOSO & \\
2021 & Wu and Huang \cite{wu2021classification}& Clinical& 38 & 2280& CNN (4)& 2Fold & \\
2021 & Huang et al. \cite{huang2021resting}& Clinical& 95 & 95& EEGNet (4) & train/valid/test split& \\
2022 & Huang et al. \cite{huang2022meg}& Clinical& 190& 33500 & ResNet (5) & 5Fold + train/val/test split & \\
2022 & Fujita et al. \cite{fujita2022abnormal}& Clinical& 180& 21000 & \begin{tabular}[t]{@{}l@{}}MLP (1) \\ MNet (12)\end{tabular} & 10Fold + holdout test set& \checkmark \\
2023 & Barik et al. \cite{barik2023functional}& Clinical& 60 & N/M & MLP (1)& 5Fold nested & \\
2024 & Anand et al. \cite{anand2024imnmagn}& Clinical& N/M& 1080000 &N/M& train/test split & \\
2024 & Achterberg et al. \cite{achterberg2024synaptic}& Clinical& 8& 2400& DNN (4)& train/valid/test split& \\
2018 & Dash et al. \cite{dash2018determining}& BCI& 4& 1635& MLP (1)& train/valid/test split& \\
2018 & Dash et al. \cite{dash2018overt}& BCI& 3& 1225& CNN (8)& 5Fold (trials)& \\
2019 & Hramov et al. \cite{hramov2019kinesthetic}& BCI& 10 & N/M & MLP& N/M& \\
2019 & Hramov et al. \cite{hramov2019meg}& BCI& 7& N/M & MLP (2)&N/M& \\
2019 & Zubarev et al. \cite{zubarev2019adaptive}& BCI& 7& 11354 & LF-CNN (3) & LOO& \\
2019 & Dash et al. \cite{dash2019automatic}& BCI& 4& 1635& CNN (5)& train/valid/test split& \\
2020 & Dash et al. \cite{dash2020decoding}& BCI& 8& 3046& \begin{tabular}[t]{@{}l@{}}AlexNet (7)\\ ResNet101 (101)\\Inception-ResNet-v2 (164)\end{tabular} & train/valid/test split& \\
2020 & Dash et al. \cite{dash2020neurovad}& BCI& 8& 3046& LSTM (3) &N/M& \\
2020 & Yeom et al. \cite{yeom2020lstm}& BCI& 9& 2160& LSTM & 5Fold & \\
2020 & Dash et al. \cite{dash2020decoding2}& BCI& 4& 1500& LSTM*& train/valid/test split& \\
2020 & Lopopolo and van den Bosch \cite{lopopolo2020part}& BCI& 15 & 1377& CNN (4)& houldout test set& \\
2021 & Ovchinnikova et al. \cite{ovchinnikova2021meg}& BCI& 32 & 422 & LF-CNN (3) & nested 4Fold in a 5Fold& \\
2022 & Fan et al. \cite{fan2022novel}& BCI& 61 & 39040 & CNN (3)& train/test split & \checkmark \\
2018 & Guo et al. \cite{guo2018stacked}& Event & 10 & 102 & SSAE (3) & 5Fold & \\
2019 & Zheng et al. \cite{zheng2019ems}& Event & 20 & 4000& CNN (8)& multiple KFold& \checkmark \\
2020 & Liu et al. \cite{liu2020novel}& Event & 20 & 150 & CNN* & KFold & \\
2021 & Zhang et al. \cite{zhang2021target}& Event & 10 & N/M & EEGNet (4) & cross-validation & \\
2022 & Hirano et al. \cite{hirano2022fully}& Event & 348& 23177 & ResNet (26) and AE & 5Fold and 10Fold & \checkmark \\
2022 & Bhanot et al. \cite{bhanot2022seizure}& Event & 15 & 11000 & N/M& 5Fold & \\
2022 & Zhao et al. \cite{zhao2022multi}& Event & 20 & 1320& DANN & Leave One Out & \\
2022 & Guo et al. \cite{guo2022transformer}& Event & 20 & 202 & Transformer & KFold & \\
2022 & Zhang et al. \cite{zhang2022efficient}& Event & 20 & 150 & \begin{tabular}[t]{@{}l@{}}CADNet\\ DendriteNet\end{tabular} & N/M& \\
2023 & Zheng et al. \cite{zheng2023artificial}& Event & 48 & N/M & CNN (5)& 7Fold & \\
2023 & Mouches et al. \cite{mouches2024time}& Event & 95 & 40662 &N/M& 10Fold& \checkmark \\
2024 & Wei et al. \cite{wei2024nested}& Event & 277& 230325&N/M& train/valid split& \\
2024 & He et al. \cite{he2024cednet}& Event & 11 & 35013 & CNN + Attention (5)& LOSO & \\
2024 & Dev et al. \cite{dev2024data}& Event & 7& 7060&N/M& train/valid/test split& \\
2024 & Hirano et al. \cite{hirano2024deep}& Event & 1782 & 60139 & 26 Layer scSE-ResNet & 5Fold subject-wise& \checkmark \\
\end{tabularx}}
 \caption{Part 2 of 4. Includes all papers categorized as 'Classification' papers and subcategorized as 'clinical', 'BCI', or 'event detection' papers.}
 \label{tab:classification2}
\end{table}
\restoregeometry

\newgeometry{textwidth=7in}
\begin{table}
 \centering
 \resizebox{.95\textwidth}{!}{
 \newcolumntype{Y}{>{\raggedright\arraybackslash}X}
\begin{tabularx}{\textwidth}{l|X|l|r|r|X|X|r@{}}
Year & Authors & Subcategory & N & trials & Architecture (depth) & Validation & Code \\
\hline
2016 & Cichy et al. \cite{cichy2016comparison}& Visual & 15 & 3540& SuperVision - CNN () & N/M &  \\
2017 & Cichy et al. \cite{cichy2017dynamics}& Visual & 15 & N/M & CNN (8)&  \begin{tabular}[t]{@{}l@{}}Bootsrapping \\ resampling \end{tabular}&  \\
2018 & Seeliger et al. \cite{seeliger2018convolutional}& Visual & 15 & 1000& VGG-S - CNN (10) & 10Fold & \\
2018 & Dima et al. \cite{dima2018spatial}& Visual & 19 & 720 & \begin{tabular}[t]{@{}l@{}}CNN (7)\\ AlexNet - CNN (8)\end{tabular}& 5Fold  & \\
2018 & Bankson et al. \cite{bankson2018temporal}& Visual & 32 & 32& VGG-F - CNN (7)& LOSO & \checkmark \\
2019 & Rajaei et al. \cite{rajaei2019beyond}& Visual & 15 & 1536& \begin{tabular}[t]{@{}l@{}}AlexNet - CNN (7)\\ HRRN (152)\end{tabular} & LOO & \checkmark\\
2019 & Kietzmann et al. \cite{kietzmann2019recurrence}& Visual & 15 & 380 & \begin{tabular}[t]{@{}l@{}}CNN \\RCNN (6)\end{tabular} &  \begin{tabular}[t]{@{}l@{}}2Fold \\ bootstrapping\end{tabular}&  \\
2020 & Giari et al. \cite{giari2020spatiotemporal}& Visual & 25 & 3525& AlexNet - CNN (5)& N/A &  \\
2022 & van Vliet et al. \cite{van2022convolutional}& Words& 15 & 560 & VGG (11) & N/A & \checkmark\\
2023 & von Seth et al. \cite{von2023recurrent}& Visual & 36 & N/M &N/M & N/M & \checkmark\\
2020 & Donhauser et al.\cite{donhauser2020two}& Speech & 11 & 77& LSTM*& 7Fold&  \\
2022 & Caucheteux and King \cite{caucheteux2022brains}& Auditory & 92 & N/M & \begin{tabular}[t]{@{}l@{}}Transformer\\ CNN (4, 8 or 12) \end{tabular}& N/A & \checkmark\\
2022 & Wingfield et al. \cite{wingfield2022similarities}& Speech & 16 & 66300 & MLP (6)& N/A &  \\
2023 & Desbordes et al. \cite{desbordes2023dimensionality}& Speech & 11 & 3630& \begin{tabular}[t]{@{}l@{}}LSTM (2)\\Transformer (11)\end{tabular}& 10Fold &  \\
2024 & Brodbeck et al. \cite{brodbeck2024recurrent}& \begin{tabular}[t]{@{}l@{}}Speech\\Recognition\end{tabular} & 18 & N/M &N/M& N/M&  \\
2024 & Lyu et al. \cite{lyu2024finding}& \begin{tabular}[t]{@{}l@{}}Speech\\Recognition\end{tabular} & 16 & 5760&N/M& N/M& \checkmark 

\end{tabularx}}
 \caption{Part 3 of 4. Includes all papers that were categorized as 'Modeling' papers.}
 \label{tab:modeling}
\end{table}
\restoregeometry

\newgeometry{textwidth=7in}
\begin{table}
 \centering
 \resizebox{.95\textwidth}{!}{
 \newcolumntype{Y}{>{\raggedright\arraybackslash}X}
\begin{tabularx}{\textwidth}{l|X|l|r|r|X|X|r@{}}
Year & Authors & Subcategory & N & Trials & Architecture (Depth) & Validation & Code \\
\hline
2019 & Dinh et al. \cite{dinh2019contextual}& SL & 1 & N/M& LSTM & train/valid split & \\
2021 & \begin{tabular}[t]{@{}l@{}}Pantazis\\ and Adler\end{tabular}& SL & N/A & N/M& CNN (5) and MLP (4) & N/A& \\
2021 & Dinh et al. \cite{dinh2021contextual}& SL & 1 & 1653 & LSTM& train/valid split& \checkmark \\
2022 & Sun et al. \cite{sun2022personalized}& SL & 29& 620256 & ResNet (4) LSTM (3) & train/test split& \\
2023 & Sun et al. \cite{sun2023deep}& SL & 29& 620256 &N/M &N/M & \checkmark \\
2023 & O’Reilly et al. \cite{o2023localized}& SL & 1 & 446& SimpleRNN (4) & Bootstrapping & \checkmark \\
2024 & \begin{tabular}[t]{@{}l@{}}Sanchez-Bort\\ et al\end{tabular}& SL & 1 & N/M& MPSS (4)&N/M & \checkmark \\
2024 & Jiao et al. \cite{jiao2024multi}& SL & 1 & N/M& CNN + Attention &N/M & \\
2024 & Yokoyama et al. \cite{yokoyama2024m}& SL & 3 & N/M& 4LCNN (4) & train/valid sets & \\
2016 & \begin{tabular}[t]{@{}l@{}}Hyvärinen\\ and Morioka\cite{hyvarinen2016unsupervised}\end{tabular}& Preprocessing & 9 & N/M& FC (4)& M/M& \\
2017 & Garg et al. \cite{garg2017automatic}& Preprocessing & 49& 980& CNN (6) & train/test split& \checkmark \\
2017 & Garg et al. \cite{garg2017using}& Preprocessing & 44& 880& CNN (9) & LOO + train/test split& \checkmark \\
2018 & Croce et al. \cite{croce2018deep}& Preprocessing & 67& 4038 & \begin{tabular}[t]{@{}l@{}}CNN (6); FC (6)\\ CNN+FC\end{tabular}& 10Fold (trials) & \checkmark \\
2018 & Hasasneh et al.\cite{hasasneh2018deep}& Preprocessing & 48& 1112 & \begin{tabular}[t]{@{}l@{}}temporal. CNN (3)\\ spatial. CNN (2)\\ sCNN + tCNN \end{tabular} & 50Fold& \\
2021 & Feng et al. \cite{feng2021automatic}& Preprocessing & 4 & 66780& GoogLeNet - CNN (144) & train/valid/test split& \\
2021 & Treacher et al. \cite{treacher2021megnet}& Preprocessing & 217 & 294& CNN (11)& 10Fold& \checkmark \\
2023 & Hamdan et al. \cite{hamdan2023reducing}& Preprocessing & N/M & N/M& AE (9)& N/M& \\
2017 & Guo et al. \cite{guo2017deep}& Method& 3 & N/M& CNN (5) & 10Fold& \\
2019 & Harper et al. \cite{harper2019exploring}& Method& 72& N/M& LRCN & train/valid/test split & \\
2021 & Abdellaoui et al. \cite{abdellaoui2021enhancing}& Method& 18& N/M& CNN + Attention& train/test split & \checkmark \\
2022 & Priya and Jayalakshmy \cite{priya2022cnn}& Method& 23& 13472& GoogLeNet &\begin{tabular}[t]{@{}l@{}}KFold with\\ holdout test set\\ \end{tabular}& \\
2023 & Gosti et al. \cite{gosti2024recurrent}& Method& 10& N/M& RHoMM - RNN (N/M) &N/M & \\
2023 & Elshafei et al. \cite{elshafei2023optimizing}& Method& 1 & 166800 & \begin{tabular}[t]{@{}l@{}}DNN (4) \\ CNN (1)\end{tabular} & train/test split& \\
2023 & Fan et al. \cite{fan2023model}& Method& 669 & 70000& \begin{tabular}[t]{@{}l@{}}LFCNN\\ VARCNN\\ HGRN\end{tabular} & train/valid split& \\
2023 & Csaky et al. \cite{csaky2023interpretable}& Method& 37& 53100&WaveNet-based &train/valid split & \checkmark \\
2023 & Zhu et al. \cite{zhu2023unsupervised}& Method & 676 & 100000 & \begin{tabular}[t]{@{}l@{}}MLP (1)\\  NICA(TCL) (3) \\ NICA(IIA) (3)\end{tabular}& N/A& \checkmark \\
2024 & Solana et al. \cite{solana2024classification}& Method& 16& N/M&\begin{tabular}[t]{@{}l@{}}ROCKET-based \\ models (3)\end{tabular}&train/valid split & \checkmark \\
2024 & Fan et al. \cite{fan2024magnetoencephalography}& Method& 250 & 29238& \begin{tabular}[t]{@{}l@{}}EEGNet\\ResNet\\ShuffleNet\end{tabular}& N/M & \checkmark \\
2024 & Gallard et al. \cite{gallard2024transforming}& Method& 44& N/M& CycleGAN&N/A & \checkmark \\
2024 & Chou et al. \cite{chou2024unified}& Method& 17& N/M& CNN (3) & train/valid/test split& \\
2024 & Csaky et al. \cite{csaky2024foundational}& Method& 15& 3540 &GPT & train/valid/test split & \\
2024 & Gideoni et al. \cite{gideoni2024non}& Method& 6 & N/M& MLP and CNN & train/valid/test split& \\
2024 & Ferrante et al. \cite{ferrante2024towards}& Method& 4 & N/M& CNN and CLIP & N/A& 
\end{tabularx}
 }
 \caption{Part 4 of 4. Includes all papers that were classified as 'Other' papers. SL is an abbreviation of source localization.}
 \label{tab:other}
\end{table}
\restoregeometry

\subsection{Types of studies}
In conducting this review, we opted to categorize the included studies into three distinct groups: (1) 'Classification', (2) 'Modeling', and (3) 'Other'. The 'Classification' category was split into four sub-groups: decoding, brain-computer interfaces (BCI), clinical applications or computer-aided diagnosis (CAD), and event detection. Studies under 'decoding' primarily leverage classification as a statistical tool to identify the involvement of specific brain features in certain tasks. The BCI subcategory includes studies that incorporate ANNs in their BCI framework. Meanwhile, the 'clinical' subcategory includes studies that utilize ANNs to enhance clinical diagnosis or prognosis, and the 'event detection' subcategory regroups studies using ANNs to continuously detect events in MEG data. Studies in the 'Modeling' category primarily focus on comparing activations across ANNs and the brain. The 'Other' category gathers the studies that do not fit the aforementioned groups and includes research utilizing ANNs for preprocessing tasks such as artifact detection and removal or source localization. The distribution across more specific subcategories is given in \ref{tab:subcat} and illustrated in figure \ref{fig:pie2}. 

\begin{table}
\centering
 \begin{tabular}{l|l|c}
 Category & Subcategory & Study \\
 \hline
 Classification 
 & Decoding & 
 \cite{kostas2019machine}, \cite{chang2021decoding}, \cite{feng2021new}, \cite{dash2019towards}, \cite{dash2020neural}, \cite{dash2019decoding}, \cite{dash2021role}, \cite{frolov2018diagnostics}, \cite{hramov2018artificial}, \cite{kim2019canet}, \cite{shu2013sparse} \\
 & & \cite{li2021inter}, \cite{wang2017towards}, \cite{yu2016encoding}, \cite{huang2019cross}, \cite{ozer2023brain}, \cite{garry2019classification}, \cite{zubarev2024robust}, \cite{pilyugina2021comparing}, \cite{bu2023magnetoencephalogram}, \\
 & & \cite{boyko2024megformer}, \cite{abdellaoui2020deep}, \cite{zhang2023decoding}, \cite{yang2024mad}, 
 \cite{csaky2023group}, \cite{engemann2022reusable}, \cite{shi2021categorizing}, \cite{yang2024neugpt}, \cite{jayalath2024brain}\\
 & BCI &
 \cite{dash2018determining}, \cite{dash2020decoding}, \cite{dash2020decoding2}, \cite{dash2020neurovad}, \cite{dash2019automatic}, \cite{dash2018overt}, \cite{hramov2019meg}, \cite{hramov2019kinesthetic}, \\
 & & \cite{fan2022novel}, \cite{ovchinnikova2021meg}, \cite{yeom2020lstm}, \cite{zubarev2019adaptive}, \cite{lopopolo2020part}, \cite{he2024cednet}\\
 & Clinical &
 \cite{aoe2019automatic}, \cite{meng2018brain}, \cite{giovannetti2021deep}, \cite{huang2021resting}, \cite{huang2022meg}, \cite{zhang2020predicting}, \cite{gu2020multi}, \cite{wu2021classification}, \cite{xu2021graph}, \cite{anand2024imnmagn}, \\
 & & \cite{achterberg2024synaptic}, \cite{fujita2022abnormal}, \cite{barik2023functional}, \cite{zhang2022efficient}, \cite{zhao2022multi}, \cite{bhanot2022seizure}, \cite{hirano2022fully}\\
 & Event detection &
 \cite{zheng2023artificial}, \cite{zheng2019ems}, \cite{guo2022transformer}, \cite{liu2020novel}, \cite{zhang2021target}, \cite{guo2018stacked}, \cite{mouches2024time}, \cite{wei2024nested}, \cite{dev2024data}\cite{hirano2024deep}\\
 Modeling 
 & Visual cortex &
 \cite{dima2018spatial}, \cite{bankson2018temporal}, \cite{cichy2016comparison},
 \cite{cichy2017dynamics},
 \cite{kietzmann2019recurrence}, \cite{seeliger2018convolutional}, \cite{giari2020spatiotemporal}, 
 \cite{van2022convolutional}
 \cite{rajaei2019beyond}, \cite{von2023recurrent}\\
 & Auditory cortex &
 \cite{donhauser2020two}, \cite{caucheteux2022brains}, \cite{desbordes2023dimensionality}, \cite{wingfield2022similarities}, \cite{brodbeck2024recurrent}, \cite{lyu2024finding}\\
 Other 
 & Preprocessing & \cite{croce2018deep}, \cite{garg2017using}, \cite{feng2021automatic}, \cite{hasasneh2018deep}, \cite{garg2017automatic}, \cite{hyvarinen2016unsupervised}, \cite{treacher2021megnet}, \cite{hamdan2023reducing}, \\
 & Methods &
 \cite{csaky2023interpretable}, \cite{abdellaoui2021enhancing}, \cite{guo2017deep}, \cite{harper2019exploring}, \cite{priya2022cnn}, \cite{fan2023model}, \cite{gosti2024recurrent}, \cite{gallard2024transforming}, \\
 & & \cite{elshafei2023optimizing}, \cite{csaky2024foundational}, \cite{solana2024classification}, \cite{fan2024magnetoencephalography}, \cite{chou2024unified}, \cite{ferrante2024towards}, \cite{gideoni2024non}, \cite{zhu2023unsupervised}\\
 & Source localization & 
 \cite{dinh2019contextual}, \cite{pantazis2021meg}, \cite{sun2022personalized}, \cite{o2023localized}, \cite{dinh2021contextual}, \cite{sun2023deep}, \cite{sanchez2024solving}, \\
 & & \cite{jiao2024multi}, \cite{ferrante2024towards}\\
 \end{tabular}
 \caption{The categories and subcategories classification for all reviewed papers. }
 \label{tab:subcat}
\end{table}

\subsection{Notes on terminology}
Some of the ML nomenclature overlaps with terminology used in M/EEG research, which can sometimes lead to confusion. In particular, the term 'epoch' in deep learning generally describes a complete pass of the training data, where the entire dataset is used once to update the model parameters. Meanwhile, in the field of M/EEG, it usually describes a segment of data (or a single trial). In addition, the term 'samples' in ML usually refers to individual examples from the dataset, which should not be confused with the use of the same term to refer to a data point in a M/EEG time series, where the parameter known as 'sampling rate' (Hz) refers to the rate at which the data is assessed, i.e., data points per second.

\section{Results}

In the following, we present a comprehensive analysis of the corpus of papers selected for this review. After a general overview highlighting key trends and patterns across various categories, we dive into a more detailed examination of the methods, data, and ANN architectures used across each of the three main categories of papers.

\begin{figure}
 \begin{minipage}{0.5\linewidth}
 \centering
 \sidesubfloat[(a)]{\includegraphics[width=.90\linewidth]{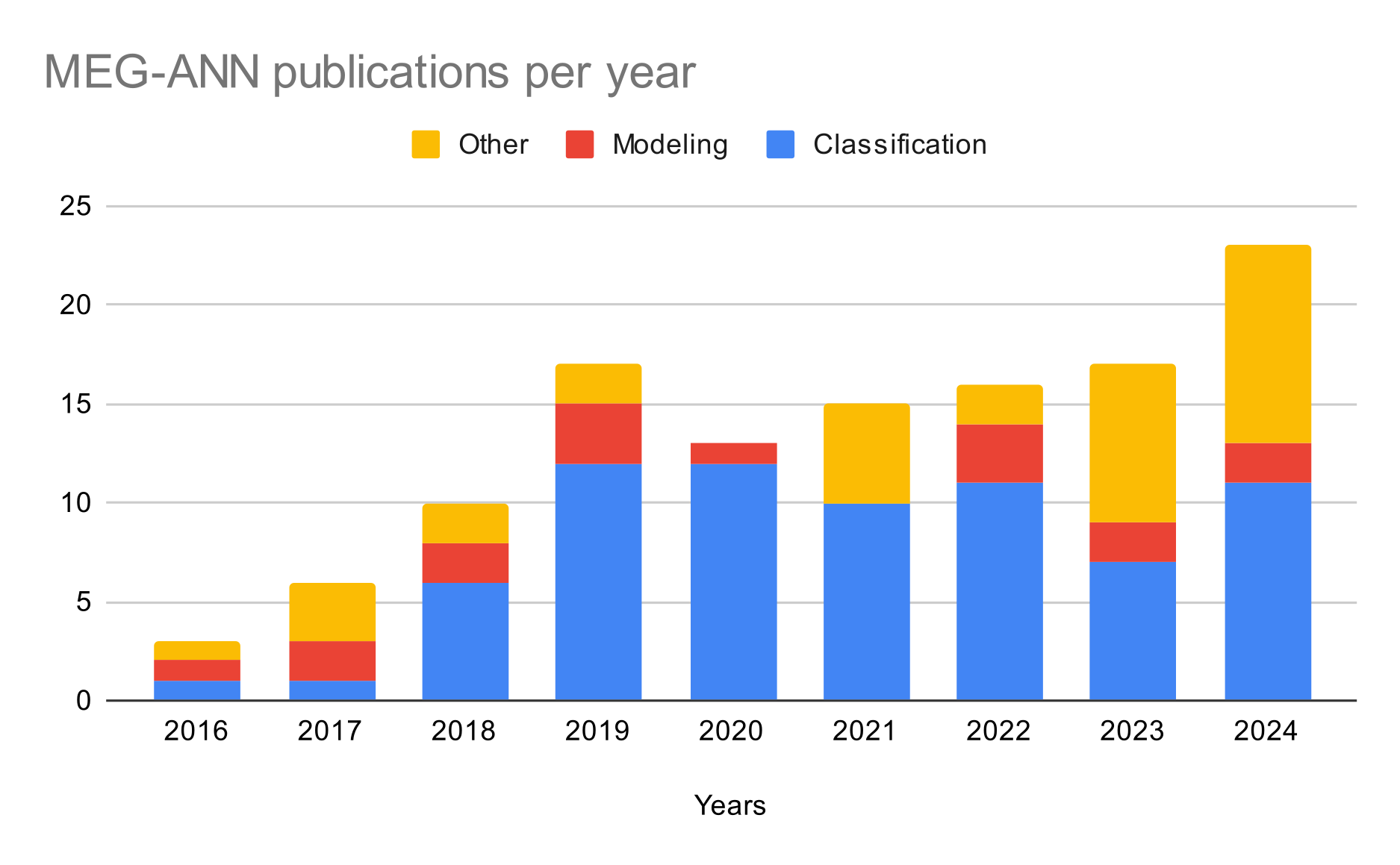}\label{fig:years}}
 \end{minipage}%
 \begin{minipage}{0.5\linewidth}
 \centering
 \sidesubfloat[(b)]{\includegraphics[width=.90\linewidth]{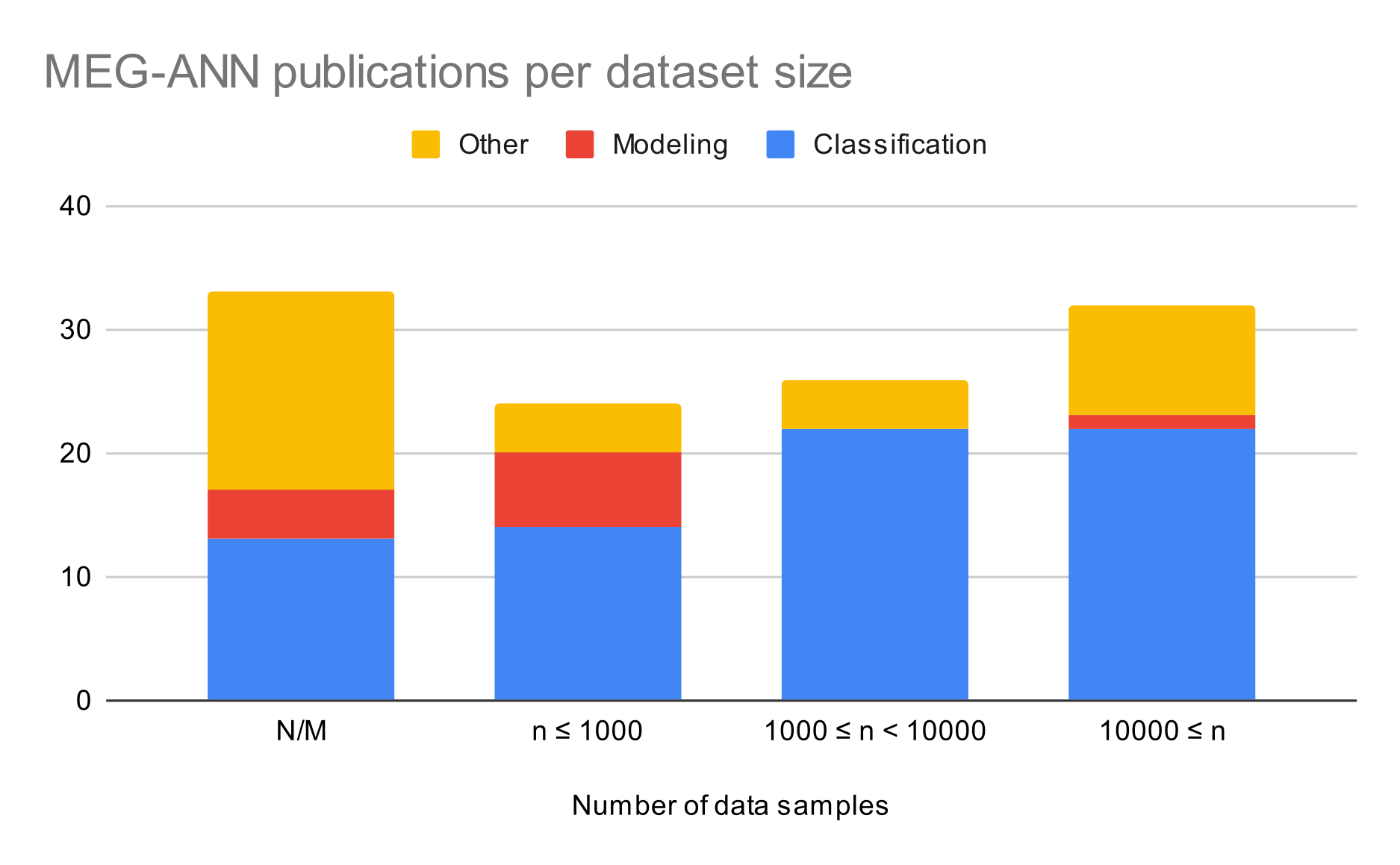}\label{fig:trials}}
 \end{minipage}
 \begin{minipage}{0.5\linewidth}
 \centering
 \sidesubfloat[(c)]{\includegraphics[width=.90\linewidth]{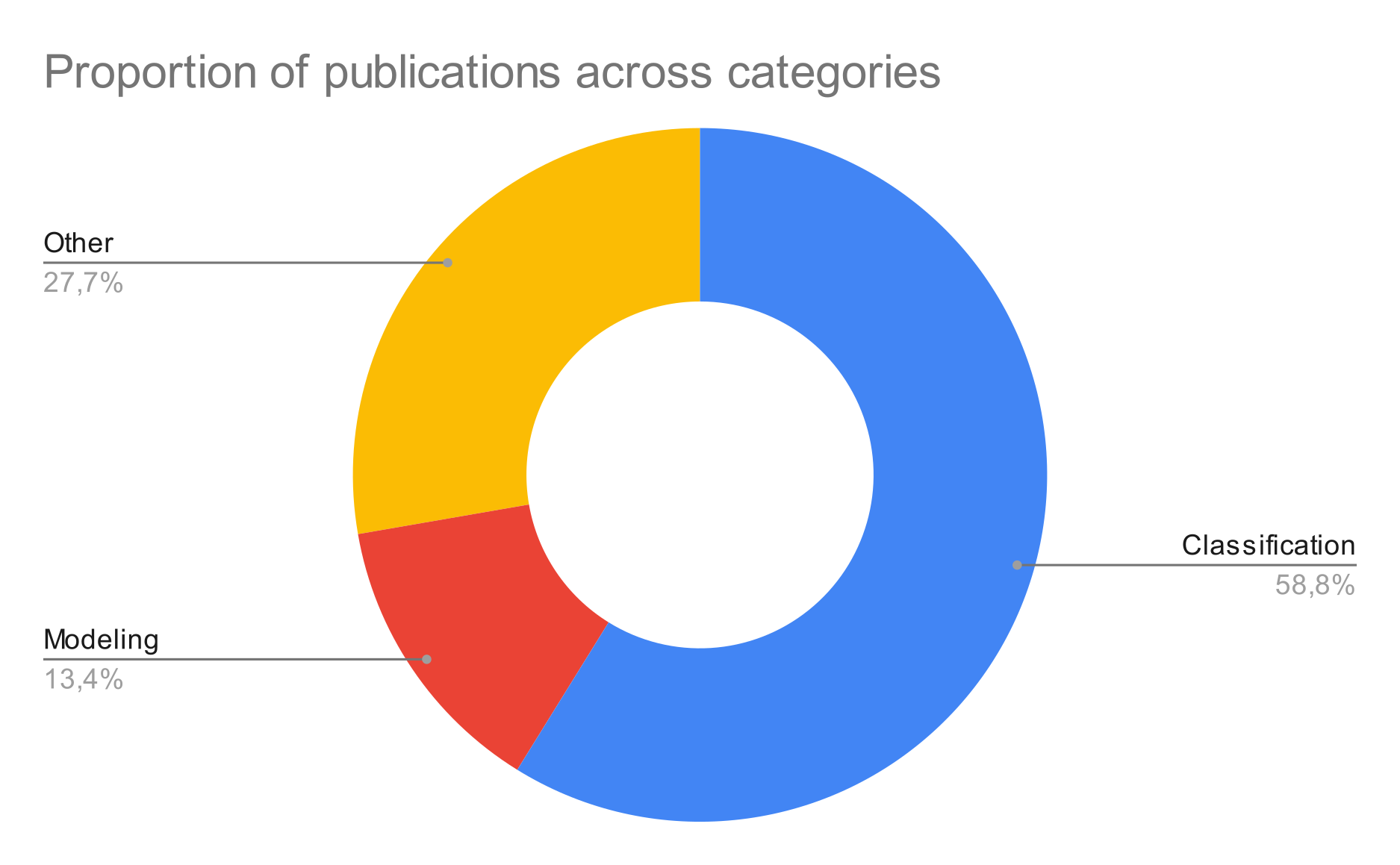}\label{fig:pie}}%
 \end{minipage}%
 \begin{minipage}{0.5\linewidth}
 \centering
 \sidesubfloat[(d)]{\includegraphics[width=.90\linewidth]{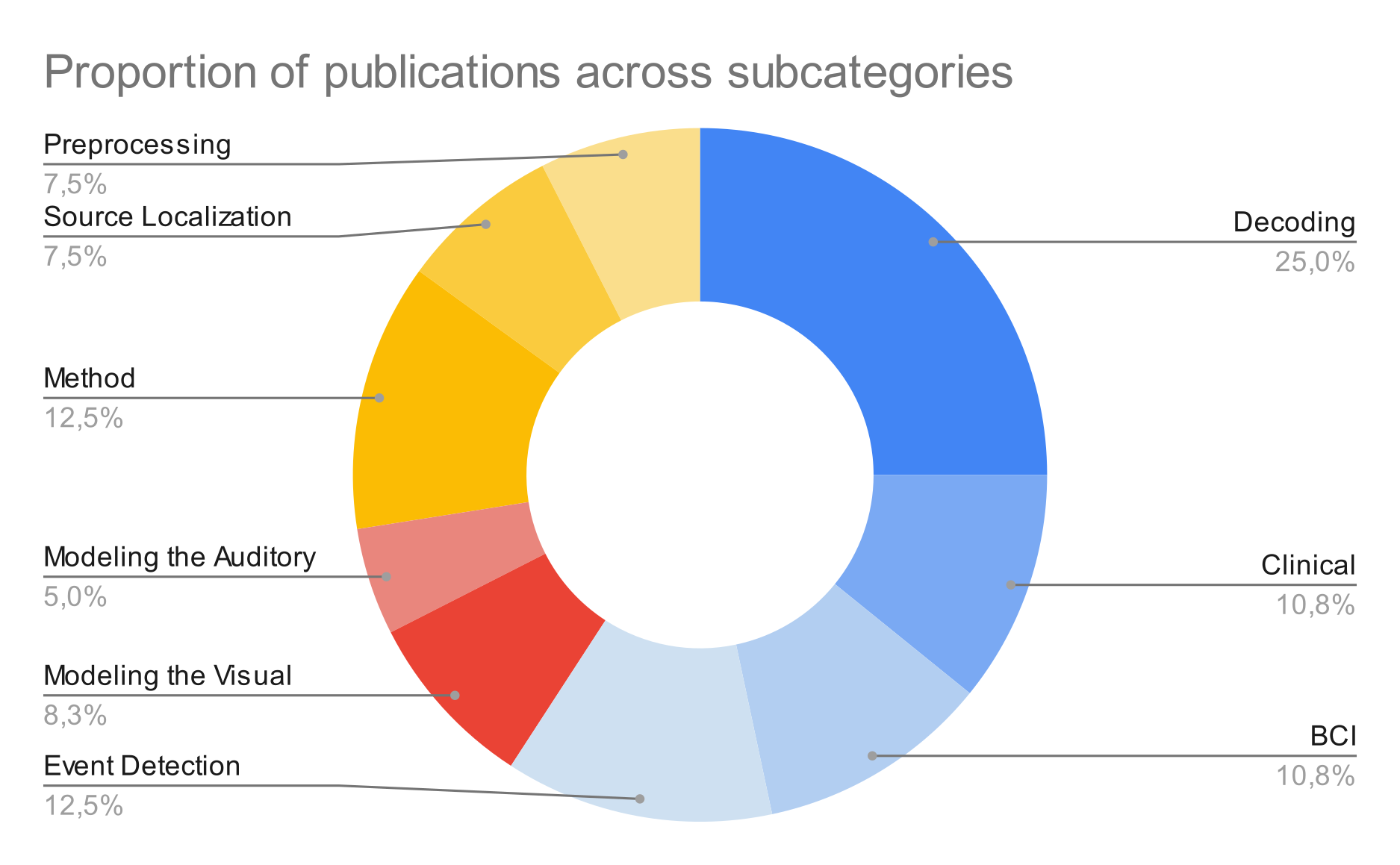}\label{fig:pie2}}%
 \end{minipage}
 \caption{Quantitative overview of the reviewed studies (n=119). (a) The evolution of the number of MEG publications using ANNs across categories (classification, modeling, and other). (b) Size of the dataset (number of samples) used for training and evaluation of the ANNs across the reviewed studies. 'N/M' accounts for the studies that did not mention the number of samples used. (c) Overall distribution across categories of all the reviewed publications. (d) Overall distribution across subcategories categories of all the reviewed publications.}
 \label{fig:bigfig}
\end{figure}

The methodology explained in the previous section (\ref{sec:methods}) led to the identification of 119 relevant studies. 
Among the papers considered, 102 were peer-reviewed, while 17 had not yet undergone peer review at the time of writing. Our categorization method resulted in the distribution of 70 papers under 'Classification', accounting for 58.8\% of the total; 16 papers were categorized as 'Modeling', making up 13.4\%; and 33 papers fell into the 'Other' category, representing 27.8\% of the study corpus included in this review (See figure \ref{fig:pie}). The category containing most of the papers is the 'Classification' category, followed by 'Other'; the category with the least amount of studies is 'Modeling'. Generally, the trend of using ANNs with MEG data has increased over the years across all categories, as depicted in figure \ref{fig:years}. A similar growth trajectory is anticipated for studies on this topic, analogous to the rapid expansion of research utilizing ANNs for EEG data analysis (\cite{roy2019deep}).

\thisfloatsetup{style=plain, subcapbesideposition=top}

\subsection{General overview}

\begin{figure}[htbp]
 \centering
 \includegraphics[width=.9\linewidth]{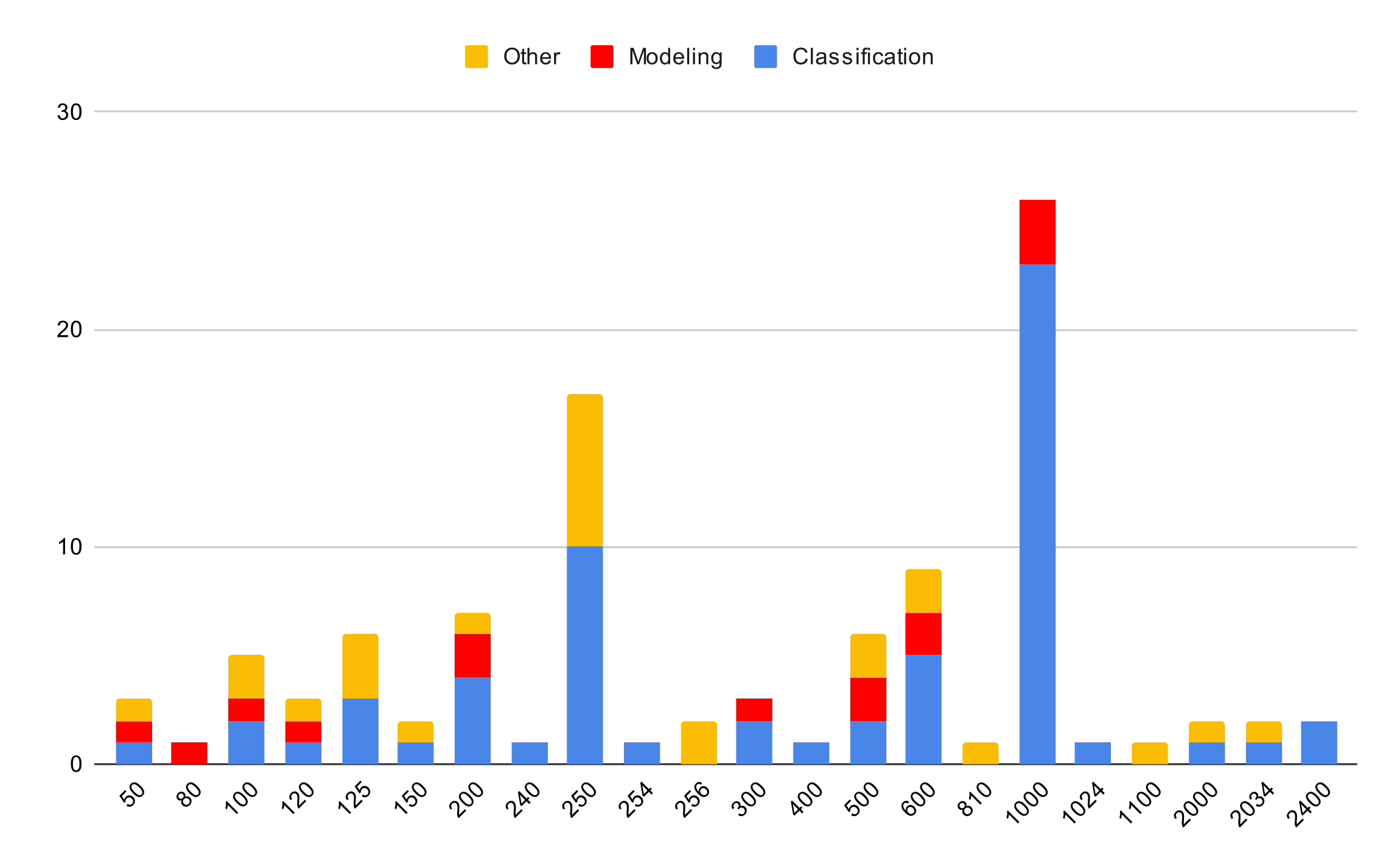}
 \caption{Distribution of sampling frequencies (in Hz) reported across the reviewed studies (N=110 reporting). The height of each bar indicates the total number of studies using a specific frequency. Stacked colors represent the main study categories (yellow: 'Other', red: 'Modeling', blue: 'Classification'). The most frequently reported rates were 1000 Hz (N=26) and 250 Hz (N=17).}
 \label{fig:samplefreq}
\end{figure}

The in-depth survey revealed several key observations about this emerging field (figure \ref{fig:bigfig}). 
Interestingly, out of the 119 selected studies for this review, 65 use less than 20 subjects, 35 use between 20 and 100 subjects, and 16 use more than 100 subjects. The remaining studies use synthetic MEG data or do not mention the number of subjects, or patients included in the study. The number of samples (number of trials or epochs) in datasets varies across a wide range, from 17 to 2304000 (see figure \ref{fig:trials}), with an average of 102247, a standard deviation of 383501, and a median of 4000. However, a few outliers significantly impact the average and standard deviation of the number of samples used in the included studies (Correcting for outliers gives an average sample number of 13246 and a standard deviation of 21147). Furthermore, the length of the data segments fed to the neural network varies between 300ms to 10s. The most commonly used sampling frequencies are 1000 Hz (26 papers) and 250 Hz (17 papers) (as illustrated by figure \ref{fig:samplefreq}), and the median across all studies was 250 Hz.

\subsection{Classification studies}
\label{sec:classif}

\subsubsection{Study aims and subcategories}

 As explained in section \ref{sec:methods}, studies in the 'Classification' category were divided into four main subcategories based on their primary goal: BCI, decoding problems, clinical applications, and event detection.

\textbf{BCI studies:} BCI studies assess the feasibility of building brain-controlled devices that rely on predicting the subject’s intentions from their neural data. BCI studies cover a wide range of approaches. Still, most consist either of offline assessments of previously recorded data or online examination of intention decoding in a closed-loop setting. BCI is one of the neuroscience domains in which ML made its earliest incursions (e.g. decoding arm movement direction and kinematics from neuronal activity in non-human primates). Of the 13 studies subcategorized as BCI, five use ANNs for motor tasks (\cite{hramov2019meg, hramov2019kinesthetic, yeom2020lstm, ovchinnikova2021meg, fan2022novel}), seven focus on speech-related BCI (\cite{dash2018determining, dash2018overt, dash2019automatic, dash2020decoding, dash2020neurovad, dash2020decoding2, lopopolo2020part}), and finally, in \cite{zubarev2019adaptive}, the authors investigate how ANNs can improve the state-of-the-art of BCI.

\textbf{Decoding studies:} Generally speaking, decoding studies aim to better understand the neural correlates of behavior and cognitive processes. This is typically accomplished by using ML or statistical inference. This approach determines the neural responses that underlie specific sensory, motor, or higher-order cognitive functions by revealing the features that exhibit the strongest separation between classes. 
In this category, eight studies use ANNs for speech decoding (\cite{wang2017towards, dash2019decoding, dash2019towards, dash2020neural, dash2021role, kostas2019machine, lopopolo2020part, boyko2024megformer}). Three studies specifically focus on using ANNs for auditory stimuli decoding (\cite{pilyugina2021comparing, feng2021new, garry2019classification}). Additionally, 10 studies use them for visual stimuli decoding (\cite{frolov2018diagnostics, hramov2018artificial, kim2019canet, li2021inter, huang2019cross, ozer2023brain, garry2019classification, csaky2023group, shi2021categorizing, zhang2023decoding}). The remaining decoding studies use ANNs for affect states decoding (\cite{yu2016encoding}), age decoding (\cite{engemann2022reusable}), hand gestures (\cite{bu2023magnetoencephalogram, zubarev2024robust}), rhythm decoding (\cite{chang2021decoding}) or text decoding (\cite{yang2024mad, yang2024neugpt, shu2013sparse}).

\textbf{Clinical studies:} we describe clinical studies, which aim to predict or detect diseases or anomalies in patients. Thirteen studies use ANNs for clinical purposes or CAD. Two focus on epilepsy-related signal properties detection (\cite{fujita2022abnormal, gu2020multi}). Two studies focus on Alzheimer's disease early detection (\cite{giovannetti2021deep}) or evolution (\cite{xu2021graph}). Three focus on neurological disease diagnostic (\cite{aoe2019automatic, achterberg2024synaptic, anand2024imnmagn}), and the remaining studies focus on schizophrenia detection (\cite{wu2021classification}), autism detection in children population (\cite{barik2023functional}), PTSD severity evaluation (\cite{zhang2020predicting}), migraine diagnostic (\cite{meng2018brain}), mild traumatic brain injury detection (\cite{huang2021resting}) or depression and bipolar disorder detection (\cite{huang2022meg}).

\textbf{Event detection studies:} Lastly, this subcategory encompasses research using ANNs to identify specific, often transient, events within continuous or segmented MEG recordings. A dominant application within the reviewed literature involves the automatic detection and sometimes localization of pathological neural events, particularly epileptic spikes or high-frequency oscillations, which are crucial biomarkers for epilepsy diagnosis and treatment planning (\cite{zheng2019ems, zheng2023artificial, wei2024nested, dev2024data, guo2018stacked, guo2022transformer, liu2020novel, mouches2024time, zhao2022multi, hirano2022fully, zhang2022efficient, bhanot2022seizure, hirano2024deep}). Other studies target different types of events, such as specific visual targets within rapid presentations (\cite{zhang2021target}). Collectively, event detection studies contributed significantly to the 'Classification' category, with epilepsy-related applications being particularly numerous (15 studies across 'Clinical' and 'Event detection' subcategories combined).

\subsubsection{The pipeline}
Across all classification studies, the typical pipeline involves using MEG recordings as input to ANNs with the goal of predicting class labels associated with sensory, motor, cognitive, or clinical states. While the precise implementation varies, most studies follow a common structure that includes preprocessing the MEG signals, optionally extracting features, and feeding the resulting data to a classifier—most often a CNN. The input of the model may consist of hand-crafted features, raw MEG signals, or a combination of both. These inputs are typically segmented into trials or epochs of fixed duration, which are then used to train and evaluate the model. The network outputs a predicted class label or probability distribution over target categories.

\begin{figure}[htbp]
 \centering
 \includegraphics[width=.9\linewidth]{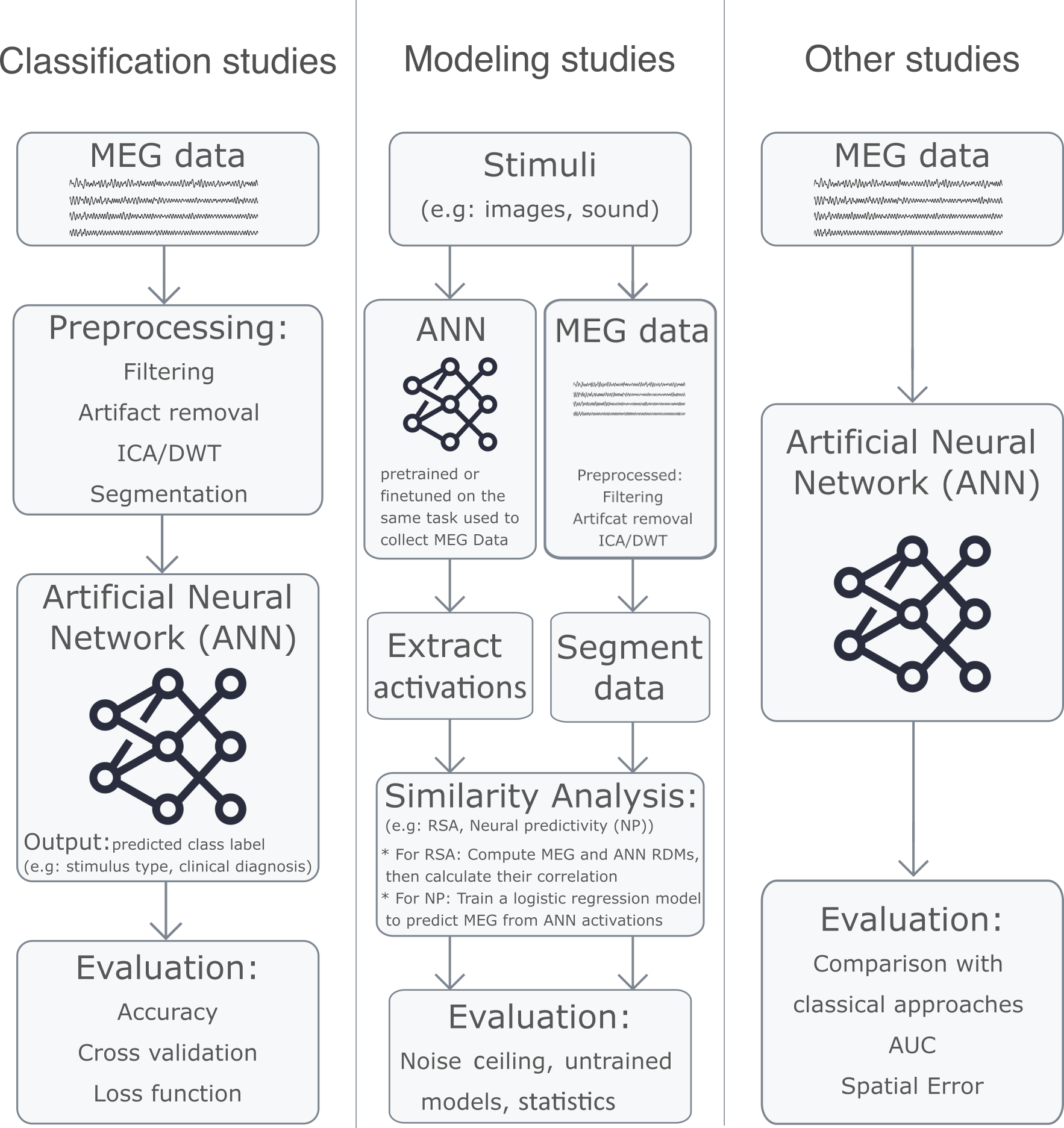}
 \caption{Representative workflows for applying ANNs to MEG data across the main application categories identified in this review. (left) 'Classification' pipeline: typically involves preprocessing MEG data, feeding it to an ANN classifier, and evaluating prediction accuracy. (center) 'Modeling' pipeline: often compares representations (activations) extracted from ANNs processing stimuli with corresponding preprocessed MEG data, using similarity analysis (e.g., RSA, neural predictivity) and evaluating against baselines. (right) 'Other' pipeline: encompasses methodological applications like preprocessing or source localization using ANNs, often evaluated by comparison with classical approaches or metrics like AUC or spatial error.}
 \label{fig:pipeline}
\end{figure}

Despite this shared pipeline, classification goals, model inputs, and preprocessing strategies differ substantially depending on the application, and no standard protocol has emerged across the 70 studies. Figure \ref{fig:pipeline} (left panel) illustrates a schematic representation of this typical 'Classification' pipeline, alongside comparable diagrams for the 'Modeling' and 'Other' application categories. 

\subsubsection{The data and preprocessing}

Participant numbers and dataset sizes varied considerably across the 70 reviewed classification studies. Due to differences in the goal and rationale, the number of subjects varied substantially (ranging from 2 \cite{wang2017towards} to 646 \cite{engemann2022reusable}). Sample sizes (trials or epochs) also varied widely; while a majority of studies reporting this information (21 out of 56) used more than 1000 samples, thirteen studies did not specify this number (see figure \ref{fig:trials} for the distribution). 

Among all classification papers, 27 studies make use of hand-crafted features. While most of these studies train the network exclusively with hand-crafted features (\cite{hramov2019kinesthetic, meng2018brain, dash2018determining, dash2020decoding, gu2020multi, dash2018overt, giovannetti2021deep, huang2021resting, xu2021graph, bhanot2022seizure, zhang2022efficient, zhang2022efficient, fujita2022abnormal, dash2019decoding, yu2016encoding, dash2020neural, dash2021role, wang2017towards, ozer2023brain, huang2022meg, achterberg2024synaptic, he2024cednet, dev2024data}), a few use a combination of hand-crafted features and raw MEG data (\cite{zhang2020predicting, feng2021new, aoe2019automatic, anand2024imnmagn}). Among the reviewed studies, spectral power and functional connectivity matrices were the two most frequently used types of hand-crafted features.

In the studies that used the original MEG data as input rather than hand-crafted features, the reported length of the segments used as samples for training and testing the ANNs varied between 125 and 2,400. This corresponds to trial durations ranging from approximately 41.67 ms to 10,000 ms, with a median of 1,063 ms across the studies that reported this information. Unfortunately, a few papers do not mention the length of the segments used as input for their neural network (\cite{hramov2018artificial, huang2019cross, chang2021decoding, anand2024imnmagn}).

 Data augmentation techniques are generally recommended in deep learning to improve model generalization and mitigate overfitting (\cite{dlbook}), but was however employed relatively infrequently in the surveyed classification studies. Specifically, only 12 out of 70 studies reported using such methods (\cite{hirano2022fully, dash2018determining, dash2019decoding, kostas2019machine, huang2022meg, abdellaoui2020deep, csaky2023group, zhang2023decoding, achterberg2024synaptic, boyko2024megformer, yang2024neugpt, hirano2024deep}). These studies use some form of linear shifting and/or sensor shuffling or mix-up regularization (\cite{carratino2022mixup}). An example of linear shifting consists of selecting a different cropping of a trial or segment around the cue in the temporal dimension. The shifted samples end up having the same length but different onsets.

Typically, across all studies in this category, the MEG data were preprocessed before being fed as input to a neural network. This preprocessing phase generally includes down-sampling, band-pass filtering, and either de-noising and/or artifact removal. Of the 70 studies in this category, 12 studies ('decoding': \cite{dash2019towards, dash2020neural, dash2019decoding, wang2017towards} 'BCI': \cite{dash2018determining, dash2020decoding, dash2020decoding2, dash2020neurovad, dash2018overt, hramov2019kinesthetic}, 'clinical': \cite{huang2022meg}, 'event detection': \cite{dev2024data}) use discrete wavelet transform to increase signal-to-noise ratio (SNR) in the data as a preprocessing step (\cite{ouahabi2013review}). Notably, these 12 studies are spread across only four distinct research groups representing a small portion of the reviewed studies. Five studies employ alternative de-noising techniques (\cite{li2021inter, csaky2023group, anand2024imnmagn, achterberg2024synaptic, zheng2023artificial}). Additionally, 16 studies use ICA to remove heart and/or eye movement artefacts (\cite{garry2019classification, bu2023magnetoencephalogram, shi2021categorizing, zubarev2019adaptive, barik2023functional, wu2021classification, zheng2019ems, zhang2021target, anand2024imnmagn, achterberg2024synaptic, zubarev2024robust, zhang2023decoding, xu2021graph, meng2018brain, giovannetti2021deep, kostas2019machine}). The remaining 38 studies do not explicitly mention de-noising or artifact removal techniques in their preprocessing pipeline.

The studies in this subcategory used MEG data sampled at frequencies between 50 Hz and 2400 Hz. Higher sampling frequencies (e.g., 1000–2400 Hz) were typically used in studies focused on decoding fine-grained temporal dynamics, such as those targeting auditory and motor-related tasks, where capturing rapid neural oscillations (e.g., in the gamma range) is critical. In contrast, lower sampling rates (e.g., 50–250 Hz) were generally the result of preprocessing choices aimed at reducing computational load or were used in studies not reliant on high-frequency information, such as certain clinical applications or feature-based pipelines. Although many studies did not report their original sampling frequency, most MEG systems (e.g., Elekta, CTF, BTi) record at native rates of 1000 Hz or above, suggesting that downsampling is commonly applied during preprocessing. Figure \ref{fig:samplefreq} displays the distribution of sampling frequencies across the reviewed studies, with 1000 Hz (26 studies of which 23 being classification studies) and 250 Hz (17 studies of wich 10 are classification studies) being the most frequently used rates, and a median of 600 Hz across all reported values. Still, eight studies omitted details about the sampling frequency, in most cases because the authors extracted features from the MEG data, making the original sampling rates less pertinent (\cite{yu2016encoding, meng2018brain, hramov2019kinesthetic, giovannetti2021deep, chang2021decoding, fujita2022abnormal, barik2023functional, dev2024data}).

\subsubsection{Network architectures}

ANNs, like any other parametric algorithm, have architectural parameters that play a key role in determining their capacity to learn from data. These parameters —including depth, number of units, and connectivity patterns— affect how well the model can extract and represent relevant features. Optimal configurations vary depending on the type of task and the nature of the input data. In this section, we provide an overview of the most commonly used architectures and their configurations across classification studies. 

The most commonly used ANNs for classification-type studies are CNNs. The second most used architecture was multi-layer perceptrons (MLP), followed by recurrent neural network (RNN) architectures. For the CNNs, the depth of the networks varied widely, between one and over 100 hidden layers (up to 164 in one case \cite{dash2020decoding}). This wide range reflects the heterogeneity in input data characteristics, task complexity, and specific CNN architectures employed across the reviewed studies, ranging from relatively shallow custom CNNs to very deep pre-existing architectures like GoogLeNet (\cite{szegedy2015going}) and ResNet (\cite{he2015deep}) variants (e.g., \cite{feng2021new, dash2020decoding}). In contrast, the papers that used MLPs essentially limit themselves to a maximum of six hidden layers. This typical depth difference often relates to parameter efficiency: the weight-sharing mechanism in CNNs generally allows for deeper architectures with high-dimensional data such as MEG recordings, whereas the large number of parameters in fully-connected MLP layers can make similar depths computationally costly and difficult to train effectively (\cite{dlbook}). 

Most of the CNN architectures used in the reported MEG classification papers comprise the input layer followed by multiple blocks, typically containing convolution layers followed by activation functions (like rectified linear unit, ReLU) and often pooling layers (like max pooling) for downsampling. At the end, one or two fully connected layers are followed by an activation function. For the convolution layers, the critical parameters are filter size, stride, padding, and number of channels. For fully connected layers, the crucial parameter is the number of units, also called the width of the layer. 

 Although an in-depth discussion of parameter initialization techniques is beyond the intended scope of this review, it is important to keep in mind that initialization methods (e.g., random, Xavier, or He initialization) are important and can profoundly influence model performance and stability (see \cite{glorot2010understanding}). Similarly, the choice of activation function (e.g., Sigmoid, Tanh, ReLU and its variants) applied after convolutional or fully connected layers is another critical design decision impacting network behavior and learning dynamics. 



\subsubsection{Training and validation techniques}

Beyond the architecture used and data (pre-)processing, details about training parameters, loss function, optimizer, and regularization techniques are essential to replicate the results of such studies. While training parameters may not have as significant an impact as architecture parameters, they remain essential and can heavily impact training time and performance. Out of the 70 classification studies, 17 do not mention training parameters. The remaining studies generally mention learning rate, batch size, loss function, optimization method, and regularization techniques. Batch sizes, regularization techniques (dropout, L1 and L2 norm weight usage), number of epochs, early stop criteria, learning rates, and momentum parameters are crucial to watch out for. While the specific training parameters used in each study are crucial for reproducibility, the high degree of variability across different experimental settings and goals makes a detailed summary within this review text impractical. Instead, to provide a resource for readers interested in specific implementations, further details on the reported parameters for individual studies can be found in the comprehensive online table (\url{https://tinyurl.com/ub3s5mr}).

Of the studies that specify the optimization algorithm utilized for training their network, 29 employed adaptive moment estimation (Adam \cite{kingma2014adam}) or one of its variants, while four used SGD. Adam, uses both the first-order and second-order moments of the gradient to adapt the learning rate, whereas SGD only uses the first-order moment. Adam is known for its faster convergence in some architectures (\cite{pan2023toward}), but it can quickly plateau after convergence (\cite{gupta2021adam}) and may perform worse than SGD in specific use cases (\cite{hassan2023effect}).

As far as the loss function is concerned, most of the surveyed studies used cross-entropy to measure the discrepancy between the predicted probabilities and the actual distribution. More specifically, it computes the loss by taking the negative log of the probability assigned to the true class, aiming to minimize this loss to improve model accuracy.

Validation techniques are essential to training any model in a classification setting, enabling the proper evaluation of the model's performance. In a setting where generalizing to new subjects is essential, such as BCI or CAD, it is crucial to exclude multiple subjects' data from the training process. This ensures the test of the trained algorithm's generalization capabilities on entirely new and unseen data. It is also essential to evaluate and fine-tune the algorithm's performances on training and validation datasets that do not share information with each other or the test set. Among the 71 studies in this category, 26 used the K-fold validation technique (K varying from two to ten), 29 used a simple train/valid/test set split, 11 used leave one out or leave P out at the subject level. Five did not mention any validation technique in their analysis pipeline. The choice between a simple train/valid/test split and cross-validation methods (like K-fold or leave-subject-out) often depends on dataset size and computational resources. While cross-validation provides more robust performance estimates and is recommended when data is scarce, as it leads to a less noisy estimate of performance (\cite{riley2020calculating}), simple splits are frequently used in deep learning when dealing with very large datasets where the computational cost of K-fold is high and a single split is deemed sufficient for stable evaluation. Reflecting this trade-off, the studies in our review using simple splits often involved substantially larger datasets compared to those employing cross-validation (see tables \ref{tab:classification} and \ref{tab:classification2}). The leave one subject out (LOSO) cross-validation strategy is particularly relevant when subject generalization is essential and the amount of available data is small. Furthermore, when using K-fold or random splits, it is crucial to consider potential subject data leakage between folds or sets, ensuring that the evaluation reflects true generalization rather than subject identification.

ANNs are powerful and can tackle a large set of problems; however, they present challenges in terms of interpretation, implementation, and tuning. In some instances, it is unnecessary to use ANNs for problems that simpler algorithms can solve. This is why it is essential when using ANNs to check whether we are using the right tool for the right problem by computing a baseline of performances that can be reached with more classical machine-learning approaches. Unlike ANNs, which can implicitly learn features, classical algorithms require careful feature selection to effectively manage the high dimensionality of MEG data. Among the papers in the 'Classification' category, 27 out of the 70 studies did not compare their ANN's performance to simpler algorithms' performances.

A critical part of decoding studies is understanding what aspect of the data allowed decoding. It is essential to find where and how information is encoded in the data to understand more about the brain. When using ML algorithms, especially ANNs, it can be hard to make sense of what is happening in the latent space. This is why using visualization tools will help with the interpretability of the network. Of the 70 papers in the 'Classification' category, only 16 included interpretation or visualization techniques applied to the network. These techniques varied widely, including methods such as visualizing activation patterns or feature maps (\cite{zeiler2013visualizing}), generating feature importance or contribution maps (e.g., using permutation feature importance \cite{breiman2001random}, saliency methods \cite{simonyan2013deep}, or additive feature attribution approaches \cite{lundberg2017unified}), analyzing performance across time or frequency, examining network-derived connectivity, and applying specific tools like gradient-weighted class activation mapping (Grad-CAM \cite{selvaraju2017grad}). This list provides illustrative examples but is not exhaustive. Further details on methods used in specific studies can be found in the supplementary online table (\url{https://tinyurl.com/ub3s5mr}). Only one of these studies was a BCI sub-categorized study (\cite{zubarev2019adaptive}), two were event detection studies (\cite{zheng2019ems, wei2024nested}), four were clinical sub-categorized studies (\cite{xu2021graph, meng2018brain, huang2022meg, barik2023functional}), and the remaining 9 were decoding studies (\cite{hramov2018artificial, kim2019canet, garry2019classification, csaky2023group, kostas2019machine, shi2021categorizing, zhang2023decoding, yang2024mad, zubarev2024robust}).

\subsubsection{Imbalanced datasets}
Despite the growing interest in using ANNs for MEG-based classification, relatively few studies explicitly address the challenge posed by imbalanced datasets. This issue is especially prevalent in clinical and event detection studies, where one class—such as a rare neurological condition or the presence of epileptic spikes—may be underrepresented in the training data. Class imbalance can significantly bias the model’s performance, as neural networks tend to favor the majority class unless corrective measures are implemented.

Among the 70 classification studies reviewed, only a handful explicitly mention strategies to deal with class imbalance. When addressed, this was most commonly achieved through the use of weighted loss functions, such as weighted cross-entropy, which assign greater penalty to errors on the minority class. Some studies adopted data-level strategies such as upsampling the minority class or synthetic data generation through techniques like synthetic minority oversampling technique (SMOTE \cite{chawla2002smote}), although these methods were rarely discussed in detail. Others relied on evaluation metrics that are more robust to imbalance, such as the F1 score, precision-recall curves, or area under the ROC curve (AUC), rather than classification accuracy. In general, however, most studies did not report class distributions or provide a rationale for their evaluation choices, making it difficult to assess the impact of imbalance on their findings.

The limited attention given to this issue is concerning, given the known sensitivity of neural networks to imbalanced data, particularly when applied to high-dimensional signals like MEG. Future work in this area would benefit from more systematic reporting of class distributions, explicit justification of performance metrics, and comparative evaluations of different strategies for mitigating imbalance (\cite{tholke2023class}). Doing so would improve the interpretability and reproducibility of results and allow for more meaningful comparisons across studies.

\subsection{Modeling studies}
\label{subsec:modeling}

\subsubsection{Study aims and subcategories} 
The idea of using AI techniques, particularly deep learning, to reverse engineer brain function has turned into a thriving topic in Neuro-AI. In the context of this review, we refer to 'Modeling' studies that use ANNs to build models of some function of the brain. In contrast to the papers that use ANNs to classify MEG data, the studies in this category do not use MEG data as input for ANNs. Instead, they primarily involve comparisons between the ANN activations and MEG recordings obtained in response to presenting the same visual or auditory stimuli to both systems (i.e. the ANN and the human brain). Therefore, the difference between the MEG-ANN modeling work and the ANN-powered MEG classification work is basic. The underlying rationale for conducting such comparisons is the hypothesis that higher similarities between the artificial and biological neural responses indicate greater functional similarities between the ANN model and the neural network mechanisms. 

Although the ANN training parameters, regularization, and validation techniques remain essential in this context for model training, they do not have the same type of impact on the results as for the classification studies discussed in the previous section. In the following, we will overview the main trends observed across the modeling studies, starting with the principle sensory processes that have been explored.

Four out of the 16 studies in this category investigated the auditory cortex (\cite{donhauser2020two, caucheteux2022brains, desbordes2023dimensionality, wingfield2022similarities}), while nine focused on the visual cortex (\cite{dima2018spatial, bankson2018temporal, cichy2016comparison, cichy2017dynamics, kietzmann2019recurrence, seeliger2018convolutional, giari2020spatiotemporal, rajaei2019beyond, von2023recurrent}). Additionally, one study aimed to model the visual word recognition in the human brain (\cite{van2022convolutional}) and two set out to model speech recognition (\cite{brodbeck2024recurrent, lyu2024finding}). 

Out of the eleven studies investigating the visual cortex (including visual word recognition), eight use representational similarity analysis (RSA \cite{kriegeskorte2008representational}) or a variant RSA-based approach, one used a method based on 'neural predictivity' (\cite{seeliger2018convolutional}), and the last one used 'profile responses' (\cite{van2022convolutional}).

Among the studies focusing on modeling the function of the auditory cortex (including speech recognition), two comparing representations of ANNs to those of the brain when presented with auditory stimuli also used a variant of RSA (\cite{wingfield2022similarities, lyu2024finding}). Three employed methods based on neural predictivity (\cite{donhauser2020two, caucheteux2022brains, brodbeck2024recurrent}). The last study employed 'dimensionality analysis' (\cite{desbordes2023dimensionality}).

\subsubsection{The pipeline} 

A schematic overview of a typical pipeline for studies in the 'Modeling' category is provided in figure \ref{fig:pipeline} (center panel). Most of the articles included in this category essentially compare neuromagnetic activity and the ANN activations generated by the same task (e.g., visual stimulus categorization).

To achieve this, MEG brain activity is typically recorded from healthy subjects as they engage in simple categorization or identification tasks involving either visual or auditory stimuli. Subsequently, these same stimuli are fed to an ANN model previously trained on a similar task and data distribution. In practice, different ANNs, such as CNNs for image stimuli and RNNs for audio stimuli, perform the same activities as the human subjects in the MEG study. This is then followed by a comparison of stimulus representation between ANNs and human brains. While some studies opt to train their ANN models from scratch, others use pre-trained models to bypass the resource-intensive training phase.

More concretely, once the models are trained, the same stimuli used for collecting brain data are fed to the ANNs and the layers' responses' are extracted. These responses are then considered as the ANNs' activity, which will then be compared to the MEG activity induced by the same set of stimuli. That said, in one study, the authors concatenated the responses from all the layers responses to create a single model-level response (\cite{cichy2016comparison}). Once the brain data are collected and the ANNs' activations have been extracted, the next step is to compare them. In the following, we will review the main two methods used across the studies we surveyed: RSA and neural predictivity. The remaining method used in this context are 'profile responses' (\cite{van2022convolutional}) and 'dimensionality analysis' (\cite{desbordes2023dimensionality}).

\textbf{Representational similarity snalysis}: RSA is an established and widely used computational method to compare patterns of neural activity across different conditions or stimuli. This is achieved by constructing and analyzing similarity matrices representing the correlation or distance between the neural responses to each pair of conditions. These similarity matrices are known as representational dissimilarity matrices (RDMs), which serve as a crucial tool in RSA by providing a quantitative measure of how neural responses differ across various experimental conditions, thus facilitating a deeper understanding of the underlying representational structures in the brain. In some studies, representational similarity matrices (RSMs) are used instead of RDMs, employing direct correlation as the similarity metric rather than $(1-\mathrm{correlation})$ for RDMs. Occasionally, decoding accuracies, an indirect measure of dissimilarity, are used instead of correlation measures (\cite{cichy2016comparison, cichy2017dynamics, desbordes2023dimensionality}). 

The methodology for computing RDMs from fMRI data is generally uniform across many studies, typically focusing on regions of interest (ROI) RDMs. In this approach, the neural activity within a specified ROI is represented by a 1D vector describing the activity of each voxel. However, there are alternative methods that do not rely on predefined ROIs. One such alternative is the searchlight RSA (sRSA) technique, which involves specifying the shape and size of searchlight regions to scan the brain. Although computationally intensive, sRSA offers the advantage of making fewer assumptions about specific brain regions.

By contrast to RSA applications to fMRI data, the high temporal resolution of MEG provides the opportunity to conduct time-resolved RSA either in the sensor or in source space. To handle this additional time dimension, six of the 11 studies using RSA consider a 1D response vector for each time point by aggregating the activity of all brain sensors or specific ROIs' voxels' responses (in the case of source space). Ultimately, they generate time-resolved RDMs, sometimes referred to as 'RDM movies' (see \cite{kietzmann2019recurrence}). This particular RSA variant was termed RDA by \cite{kietzmann2019recurrence}. Among the remaining studies employing RSA, two used a temporal variant from searchlight RSA, termed spatiotemporal sRSA (ssRSA \cite{wingfield2022similarities}). Instead of computing RDMs for every time point, the authors of these studies use time segments of a fixed length (25 ms for \cite{wingfield2022similarities} and 16 ms for \cite{dima2018spatial}). These studies combined this approach with searchlight RSA to construct multiple RDMs for different regions using a specific time interval. The final studies represented a combination of both approaches (\cite{giari2020spatiotemporal, von2023recurrent, lyu2024finding}).

The eight studies computed the RDMs for the ANNs, one RDM per layer, with an additional RDM per model in one study. 

Finally, a similarity score is computed by correlating the RDMs of MEG data with those of ANNs. The objective is to evaluate the similarity between the geometrical representations , defined as neural activity patterns organized into a geometric space according to the distances among these patterns, of MEG and ANNs. This comparison is performed across both time (how MEG neural patterns evolve and align temporally with ANN activations) and space (how spatial patterns from MEG sensors or sources relate to ANN activations). Such comparison reveals whether and when ANNs produce neural representations resembling those of the human brain. For more information on RSA, see \cite{diedrichsen2017representational, nili2014toolbox, kriegeskorte2008representational}.

\textbf{Neural predictivity}: As an alternative approach to RSA, neural predictivity essentially involves using an ANNs' layer activity to predict brain activity. This prediction is typically achieved using an ML algorithm, such as linear regression or a support vector machine. Some subjects are used to train the ML model, while the remaining subjects are kept aside as a test set. The algorithms' performances are evaluated by comparing the predicted brain activity with the actual recorded brain activity, using correlation coefficients or mean squared error. We refer the readers to \cite{van2017primer} for further details.

\textbf{Multivariate pattern analysis (MVPA)}: Nine out of the 16 studies surveyed in this category begin with an MVPA analysis (\cite{dima2018spatial, bankson2018temporal, cichy2016comparison, cichy2017dynamics, giari2020spatiotemporal, van2022convolutional, rajaei2019beyond, desbordes2023dimensionality}). 
MVPA is a statistical approach in neuroimaging that uses standard machine learning algorithms, such as LDA and SVM, to analyze and interpret patterns of brain activity across multiple voxels or sensors, thereby facilitating the identification of cognitive states from thorough and varied neural data. But why do many studies in this category begin by using MVPA? The primary objective of conducting MVPA analysis before initiating the similarity analysis between MEG and ANN is to help identify patterns and features in the data most relevant for distinguishing between different mental states or stimuli. Applying MVPA first effectively reduces the dimensionality of the MEG data. It brings focus to the most informative features, which can then be used to compare stimulus representations across artificial and biological networks.

It is helpful to note here that out of the studies which used MVPA, three actually used MVPA results (i.e. decoding accuracy) to build similarity measures (\cite{cichy2016comparison, cichy2017dynamics, desbordes2023dimensionality}). Furthermore, some studies described the entire process of classification followed by similarity analysis as MVPA.

\subsubsection{The data and preprocessing}

Across the 16 modeling studies reviewed, the number of participants ranged from 11 to 92 (median = 15), and the number of trials or samples ranged from 77 to 66,300 (median = 3,585). However, five studies did not report the number of samples, and none reported the number of trials used for model training explicitly, as these studies did not train ANNs on MEG data.

It is important to note that, unlike classification studies where MEG data serve as input to ANNs, modeling studies in this review did not train ANNs using MEG signals. Instead, it primarily compares ANN activations with MEG recordings, observing responses to the same visual or auditory stimuli presented separately to both systems. For instance, in the study by Kietzmann et al. (\cite{kietzmann2019recurrence}), the input data consists of images from different categories (animate or inanimate objects, faces, etc). Although MEG data are not used as inputs to the ANNs in these studies, they are still used to compare the representations in brain signals and the latent space variables of ANNs. 

All MEG data collected for the studies in this category were preprocessed using a similar procedure, which in principle consists of a band-pass filter, an artifact removal and/or correction technique, and data down-sampling. The band-pass filtering's lower and upper cutoff frequencies are 0.03 or 0.1 Hz to 300 or 330 Hz. However, there seems to be no discernible pattern explaining the sampling frequencies (figure \ref{fig:samplefreq}), the de-noising, or artifact removal techniques used (see table \ref{tab:modeling}). Among the 10 studies using RSA, five have used source reconstruction before computing RDMs (\cite{dima2018spatial, kietzmann2019recurrence, wingfield2022similarities, van2022convolutional, von2023recurrent}). It is worth noting that only two studies explored the frequency domain of MEG data by investigating the similarities in one or more frequency bands (\cite{desbordes2023dimensionality, caucheteux2022brains}).

\subsubsection{Network architectures}

The choice of ANN architecture is a particularly relevant aspect of the ANN-MEG modeling studies because the ultimate goal here is to assess and interpret similarities and discrepancies between the way information is processed in biological and artificial networks. As a general rule, the articles that focus on the visual cortex use a CNN architecture trained on image classification. Out of these studies, one incorporates lateral connections between layers, in addition to the typical top-down connections, introducing temporal relationships in their ANN (\cite{kietzmann2019recurrence}). In this study, the ANN is trained to learn how to recreate MEG data RDMs from the stimuli image displayed to the subject. This is achieved through changing the ANN's objective function. About half of the studies interested in the auditory cortex use LSTM (\cite{desbordes2023dimensionality, donhauser2020two}), the remaining studies use transformer network and a CNN (\cite{caucheteux2022brains}), an MLP architecture (\cite{wingfield2022similarities}) or the BERT (\cite{koroteev2021bert}) model (\cite{lyu2024finding}). Finally, in \cite{van2022convolutional}, the authors investigated the neural responses for a visual word recognition task by comparing the visual cortex's response patterns to those of CNN architectures.

We found that while a few studies in this category train the networks from scratch, the majority (nine out of the 16) use pre-trained architectures from well-established types of networks, including VGG variants such as VGG-S (streamlined), VGG-F (fast), and VGG-11 (11 layers), which excel in image recognition. Other popular architectures include AlexNet, known for its effectiveness in image classification; CORnet-S, a brain-inspired model for predicting neural responses (\cite{kubilius2019brain}); and BERT, a transformer-based model widely used in natural language processing.
 
\subsubsection{Training and validation techniques}

In modeling studies, the emphasis is placed less on predictive performance and more on the alignment between ANN representations and brain responses. As a result, training and validation procedures are not always detailed with the same rigor as in classification-focused work. Still, when custom architectures are trained, typical practices include specifying the optimization algorithm, loss function, and stopping criteria, though many studies using pre-trained networks omit these details entirely.

Among the studies that reported their training procedures, most adopted standard deep learning practices such as stochastic optimization and loss minimization over a supervised objective. The Adam optimizer was frequently employed due to its computational efficiency and robustness to noisy gradients, although few papers provided full training specifications. In modeling studies, ANNs are typically pretrained or fine-tuned on tasks unrelated to MEG data though these tasks are often similar to the experimental conditions used to collect the MEG data (e.g., object recognition, language processing). Training and validation procedures focus on the model’s primary objective (e.g., classification or language modeling), while representational alignment with MEG signals is assessed in a separate analysis using similarity metrics such as RSA. These approaches differ fundamentally from classification pipelines, as MEG data is not used to train or validate the model itself, but rather to evaluate how well its internal representations reflect brain activity.

\subsubsection{Estimating performance baselines}

Assessing baseline performance is crucial for benchmarking and evaluating the significance of the observed similarities. Essentially, by establishing what can be expected by chance, or alternatively, what is maximally achievable given the noise in the data, we can confidently assert whether the similarities measured by RSA reflect genuine and meaningful correspondences between the representations in artificial neural networks and those observed in biological neural activity. In the following, we review the main methods of baseline performance used in the reviewed papers, focusing on noise ceilings (NC) and assessment of the untrained model's performance.

\textbf{Noise ceiling}, or the shared-response model (SRM) comparison, refers to the theoretical upper limit or maximum level of similarity that can be achieved in the absence of measurement noise or variability in brain data. In other words, a noise ceiling provides an estimate of the best performance any model can achieve given the noise in the data. As such, it serves as a benchmark for the observed similarities between ANNs and the brain, indicating the extent to which individual subjects' brain responses can be explained with a model-free approach. This concept often acts as a proxy for signal-to-noise ratio analysis, aiding researchers in interpreting the significance and reliability of their observed similarities by considering the inherent noise or variability in the data. As is common in RSA, the upper noise ceiling is estimated as the mean correlation between the group-average RSM and each participant-specific RSM. The lower noise ceiling is estimated as the mean correlation between the group-average RSM and each participant-specific RSM while iteratively excluding a given participant from the group-average. Some studies report both upper and lower noise ceilings. Although it is an essential measure, only five of the surveyed studies reported it. Among them, four used RSA as a similarity method (\cite{dima2018spatial, bankson2018temporal, kietzmann2019recurrence, von2023recurrent}), one used neural predictivity (\cite{caucheteux2022brains}) and the last used profile responses (\cite{van2022convolutional}). For detailed instructions on how to compute the noise ceiling, refer to \cite{bankson2018temporal}.

\textbf{Untrained models performance as baseline:}
In RSA studies, a common approach to assess performance is to compare the results obtained with the trained ANN model to those obtained with an untrained ANN model. Some RSA-based studies (e.g. \cite{millet2021inductive}) have found that even randomly initialized models can exhibit some similarity to the neural representation. Such an intrinsic similarity could result from built-in properties of specific models' architecture (e.g., convolutional layers resembling the visual cortex). However, most RSA studies make implicit assumptions (or at least have some expectations) that training ANNs would lead to enhanced ANN-brain similarities. In this context, comparisons of RSA results obtained with trained and untrained models can be very informative. Interestingly, out of the sixteen studies we found in this category, only three contrast their modeling results with those of a random model (\cite{cichy2016comparison, van2022convolutional, caucheteux2022brains}).

While most studies focus on a single architecture, some include multiple ANN models. By conducting RSA analyses between the brain data and each one of the models, such studies can pick up the specific network architecture and training properties that increase the similarities between the ANN and biological responses. In \cite{dima2018spatial}, the authors employed a feature-based model where RDMs are computed based on the features extracted from the stimuli. In \cite{cichy2016comparison}, the authors added a model trained on noise and an unecological model where images have been assigned random labels. Furthermore, in \cite{bankson2018temporal}, the authors compared their similarities with those obtained by a semantic model. In \cite{von2023recurrent}, the authors use semantic models that measure semantic similarity between objects, i.e., the degree of resemblance in meaning between two pieces of text, such as words or sentences. In addition, a new trend consists of training the same ANN on several related tasks (or use different objective functions) to investigate how each training goal shapes the learned representations (\cite{conwell2024large, kanwisher2023using, dobs2022brain}).

\subsection{Other studies}
\label{sec:other}

\subsubsection{Study aims and subcategories}
The final group of studies reviewed includes all MEG-related research involving artificial neural networks that do not fall neatly into either the 'Classification' or 'Modeling' categories. These works span a broad range of objectives and methodologies, including preprocessing pipelines, source localization, and the development of novel methods or architectures. Despite their diversity, these studies share a common goal: improving the utility, interpretability, and methodological foundations of MEG analysis using artificial neural networks.
This category includes 33 studies and can be divided into three broad subcategories. 

The first subcategory focuses on preprocessing techniques \cite{croce2018deep, garg2017using, feng2021automatic, hasasneh2018deep, garg2017automatic, hyvarinen2016unsupervised, treacher2021megnet, hamdan2023reducing}, proposing ANN-based tools to enhance signal quality by improving artifact detection, SNR, or visualization of neural dynamics. These studies often frame their contributions as supplements to or replacements for traditional techniques like ICA or wavelet filtering.

The second subcategory encompasses studies addressing the source localization problem, also known as the MEG inverse problem \cite{dinh2019contextual, pantazis2021meg, sun2022personalized, o2023localized, dinh2021contextual, sanchez2024solving, jiao2024multi, sun2023deep, yokoyama2024m}. These studies typically train ANNs, often using simulated MEG data, to predict source locations and commonly evaluate performance by comparison to established inverse modeling techniques such as minimum norm estimate (MNE) \cite{hamalainen1994interpreting}, Beamforming \cite{van1988beamforming}, or sLORETA \cite{pascual2002standardized}.

The final group includes methods-oriented studies \cite{csaky2023interpretable, abdellaoui2021enhancing, guo2017deep, harper2019exploring, priya2022cnn, fan2023model, gosti2024recurrent, gallard2024transforming, elshafei2023optimizing, csaky2024foundational, solana2024classification, fan2024magnetoencephalography, chou2024unified, ferrante2024towards, gideoni2024non, zhu2023unsupervised} that introduce new architectures, training strategies, or analysis frameworks designed to advance ANN-based MEG research. These studies often overlap with classification tasks but are primarily methodological in focus, aiming to improve neural representation learning, domain adaptation, or multimodal alignment rather than solving specific neuroscientific questions.

Although the aims of the studies in this category are heterogeneous, they collectively demonstrate the potential of ANNs to enhance the entire MEG analysis pipeline—from raw data handling to high-level inference. In the following sections, we outline the typical processing workflows used across these studies and examine their data, architectures, training strategies, and evaluation frameworks.


\subsubsection{The pipeline}
Given the diverse aims of the studies in this category, the pipelines vary considerably depending on whether the study focuses on preprocessing, source localization, or methodological innovation. Nonetheless, each subcategory follows a relatively coherent structure that reflects its specific objectives. Figure \ref{fig:pipeline} (right panel) illustrates a generalized pipeline applicable to many 'Other'-categorized studies.

In preprocessing-focused studies, the pipeline generally begins with raw MEG data acquisition followed by segmentation and optional band-pass filtering. These signals are then passed to an ANN architecture trained to detect and correct artifacts, enhance signal quality, or extract meaningful representations. For instance, some studies used deep convolutional networks to classify time-series segments as clean or contaminated by artifacts such as blinks, saccades, or heartbeats \cite{treacher2021megnet}, while others proposed autoencoder-based approaches to denoise the data and simultaneously improve interpretability \cite{hamdan2023reducing}.

In source localization studies, the pipeline often begins with simulated dipolar sources projected to sensor space using forward models. These simulated signals, sometimes mixed with noise at different levels, serve as input for training ANNs tasked with predicting the original cortical source locations. Once trained, these models are validated either on additional simulated datasets or real MEG recordings. The ANN output typically includes spatial maps or coordinate predictions that are compared against known ground truths or the output of classical methods like MNE, Beamformers, or sLORETA \cite{dinh2019contextual, pantazis2021meg, sun2022personalized, o2023localized, dinh2021contextual, sanchez2024solving, jiao2024multi, sun2023deep, yokoyama2024m}.

Methods-focused studies typically design and validate new architectural or learning frameworks, often repurposing existing MEG datasets for benchmarking. The pipeline usually involves adapting a neural network for a specific task—such as decoding, temporal forecasting, or multimodal alignment—and then comparing its performance to standard models or techniques. In these studies, the output may be a classification, a reconstructed signal, a latent representation, or a learned alignment between modalities \cite{csaky2023interpretable, abdellaoui2021enhancing, guo2017deep, harper2019exploring, priya2022cnn, fan2023model, gosti2024recurrent, gallard2024transforming, elshafei2023optimizing, csaky2024foundational, solana2024classification, fan2024magnetoencephalography, chou2024unified, ferrante2024towards, gideoni2024non, zhu2023unsupervised}. In some cases, these architectures are trained to generalize across datasets or tasks, with evaluation metrics designed to test their robustness, transferability, or explanatory power.

Despite the variability in input-output goals across these pipelines, what unites them is the emphasis on enhancing MEG data processing through neural network-driven components. Whether applied to raw data, source estimation, or methodological refinement, ANNs are increasingly used as flexible tools capable of improving the accuracy, interpretability, or automation of MEG workflows.

\subsubsection{The data and preprocessing}
The types of data used in this category are as varied as the study objectives, encompassing raw MEG recordings, independent components (ICs), connectivity graphs, and even anatomical MRI when required for source modeling. Most preprocessing and method studies worked directly with raw or minimally processed MEG signals, while source localization studies typically relied on simulated data for training and anatomical priors for real-world testing.

Across the 33 studies in this group, the number of trials used ranged widely, from as few as 294 to over 620,000, and the number of subjects varied between three and 676. However, nine studies did not report the size of their dataset, and two omitted the number of participants. When MEG data were used, downsampling was common. In most cases, the data were resampled to around 250 Hz (see figure \ref{fig:samplefreq}, although some studies retained higher sampling rates—such as 2000 Hz \cite{yokoyama2024m} or 2034 Hz \cite{abdellaoui2021enhancing}—to preserve fine-grained temporal information. Preprocessing routines generally included band-pass filtering and artifact removal, although detailed procedures were not always specified.

In preprocessing-oriented studies, the raw signals were typically cleaned using either manual ICA, automatic artifact detection algorithms, or ANN-based classifiers. For example, one study introduced time contrastive learning (TCL) as an unsupervised alternative to traditional ICA for separating neural from non-neural sources \cite{hyvarinen2016unsupervised}, while another employed a hybrid deep learning architecture combining 1D and 2D CNNs to detect and remove a wide range of artifacts, including eye and cardiac activity \cite{treacher2021megnet}. Others used denoising autoencoders to enhance SNR and support visualization of the signal structure \cite{hamdan2023reducing}.

In the source localization subcategory, seven studies explicitly used simulated datasets for training, leveraging ground truth source positions to supervise the learning process \cite{o2023localized, sun2022personalized, pantazis2021meg, sanchez2024solving, hamdan2023reducing, sun2023deep, yang2024mad}. These datasets were often augmented with varying levels of noise to ensure robustness. After training, the models were validated on real MEG recordings, with or without coregistered MRI data. Anatomical information was used to refine spatial accuracy, either by constraining predictions to the cortical surface or by integrating MRI-based head models into the forward projection process.

While some preprocessing and methods papers reused data from previous decoding studies, others evaluated their approach across multiple datasets to test generalizability. Despite this opportunity, only nine of the 16 methods-focused studies applied their model to more than one dataset or task type \cite{fan2023model, abdellaoui2021enhancing, harper2019exploring, gallard2024transforming, csaky2024foundational, chou2024unified, fan2024magnetoencephalography, gideoni2024non, ferrante2024towards}. This highlights the ongoing challenge of establishing robust benchmarks in ANN-based MEG methodology research.

\subsubsection{Network architectures}
The diversity of goals in this category is reflected in the wide variety of network architectures employed. While CNNs dominated preprocessing and methodological studies, source localization studies tended to rely on architectures with temporal modeling capabilities such as recurrent networks.

In preprocessing and methods papers, CNN-based architectures were the most commonly used. These included both standard 1D or 2D CNNs \cite{garg2017using, treacher2021megnet, hasasneh2018deep, croce2018deep, garg2017automatic, feng2021automatic, priya2022cnn, fan2023model, guo2017deep, pantazis2021meg, elshafei2023optimizing, fan2024magnetoencephalography, chou2024unified, yokoyama2024m, ferrante2024towards, gideoni2024non} and more specialized variants such as attention-augmented CNNs \cite{abdellaoui2021enhancing, jiao2024multi} or recurrent CNNs \cite{harper2019exploring}. The choice of CNNs is often motivated by their ability to capture spatiotemporal patterns in MEG signals, and in some cases, to operate directly on time-frequency representations or connectivity matrices. Autoencoders were also employed, especially in unsupervised preprocessing pipelines aiming to denoise or reconstruct input signals \cite{hamdan2023reducing}.

In source localization studies, architectures that explicitly modeled temporal dynamics were more common. These included long short-term memory (LSTM) networks and other types of RNNs, which are well-suited to tracking changes in MEG activity over time. Some studies also explored ResNet-based models \cite{sun2023deep} or more novel frameworks such as the multiple penalized state space (MPSS) architecture, designed to integrate spatial and temporal priors \cite{sanchez2024solving}. A few studies combined recurrent and convolutional elements to capture both local and distributed features in the MEG signals.

Method-focused studies, in contrast, emphasized innovation in architecture or training paradigms. For instance, one study introduced a contrastive learning framework to align neural and visual representations across modalities \cite{ferrante2024towards}, while others trained foundational models for MEG forecasting and decoding, achieving state-of-the-art performance across several benchmarks \cite{csaky2024foundational}. Another novel direction was the use of generative adversarial networks, such as CycleGAN, to enhance signal translation between domains or conditions \cite{gallard2024transforming}.

Across all subcategories, the architectures were tailored to specific goals—whether improving signal quality, solving inverse problems, or developing transferable decoding systems. However, these studies also highlighted that ANN architecture alone is rarely sufficient: success often depends on the thoughtful combination of network design, data preparation, and evaluation strategies.

\subsubsection{Training and validation techniques}
Training and validation procedures varied across studies in this category, depending largely on the objective and data type used. Nonetheless, most studies acknowledged the importance of robust evaluation to assess generalization performance, especially in the context of methodological innovation or when working with small datasets.

In preprocessing and method-oriented studies, training typically followed standard supervised or unsupervised learning procedures. When labeled data were available, supervised learning was used with optimization algorithms such as Adam or SGD, often in combination with dropout, batch normalization, and early stopping to prevent overfitting. Several studies also employed data augmentation strategies to improve generalizability, including sensor shuffling, temporal jittering, or synthetic data generation \cite{abdellaoui2021enhancing, treacher2021megnet, hasasneh2018deep, gallard2024transforming}. In studies leveraging unsupervised approaches like autoencoders or contrastive learning, training aimed to minimize reconstruction loss or maximize representation alignment rather than task-specific accuracy.

Validation strategies in these studies included simple train/validation/test splits, K-fold cross-validation, and LOSO schemes. LOSO was most commonly used when generalizing across participants was crucial or when datasets were small. K-fold and bootstrap resampling methods were employed to ensure robustness, particularly when evaluating architectural innovations across multiple configurations. In method papers proposing general-purpose architectures, performance was often tested across different tasks or datasets to demonstrate transferability.

In source localization studies, validation was often carried out in two stages. First, models were trained and evaluated on simulated data with known source locations and varying levels of noise, allowing precise quantification of localization error using spatial accuracy metrics or distance from the ground truth. Second, the models were applied to real MEG data, and their outputs were compared to those produced by traditional methods such as MNE, Beamforming, or sLORETA \cite{o2023localized, sun2022personalized, pantazis2021meg, sanchez2024solving, hamdan2023reducing, sun2023deep, yang2024mad}.

Overall, validation was a key element in determining the reliability and potential applicability of the proposed networks. However, the breadth of tasks and data types makes direct comparisons across studies challenging. Greater consistency in reporting training durations, regularization strategies, and evaluation metrics would be beneficial for future work seeking to benchmark ANN-based pipelines in MEG analysis.

\subsubsection{Comparison to classical approaches}
A distinguishing feature of many studies in this category is their explicit comparison between ANN-based methods and more traditional MEG analysis techniques. These comparisons were critical for justifying the adoption of neural networks, particularly in areas where classical approaches remain well-established, such as source localization or artifact correction.

In the source localization subcategory, ANN models were frequently benchmarked against established inverse solutions, including minimum norm estimate (MNE) \cite{hamalainen1994interpreting}, Beamformer \cite{van1988beamforming}, and variations of LORETA (e.g., eLORETA, sLORETA) \cite{pascual2002standardized}. Metrics used for comparison included localization error, area under the curve (AUC), and spatial dispersion. Several studies showed that ANN-based models —particularly those trained on simulated data— were capable of achieving localization accuracy that matched or exceeded classical methods, especially in low signal-to-noise regimes or when dealing with complex source configurations \cite{o2023localized, sun2022personalized, pantazis2021meg, sanchez2024solving, hamdan2023reducing, sun2023deep, yang2024mad}. Recent contributions, such as ConvDip \cite{hecker2021convdip}, demonstrated that convolutional architectures can provide competitive performance while being more robust to noise and variability in source configurations.

In preprocessing-focused studies, the goal was often to improve artifact detection and signal enhancement beyond what traditional ICA or heuristic thresholding could achieve. For instance, one study replaced ICA with time contrastive learning (TCL), an unsupervised deep learning framework that automatically separated neural from non-neural sources \cite{hyvarinen2016unsupervised}. Another employed a deep CNN-based artifact classifier that outperformed manual rejection and conventional statistical approaches in identifying blinks, heartbeats, and other common contaminants \cite{treacher2021megnet}.

In the methods subcategory, comparisons to classical baselines were essential for demonstrating the added value of architectural or algorithmic innovations. Thirteen studies explicitly compared their ANN models to simpler or more common alternatives \cite{croce2018deep, garg2017using, dinh2021contextual, pantazis2021meg, zhu2023unsupervised, o2023localized, sun2022personalized, solana2024classification, csaky2024foundational, fan2024magnetoencephalography, chou2024unified, jiao2024multi, yokoyama2024m}. These comparisons often focused on decoding accuracy, robustness to inter-subject variability, and computational efficiency. Some studies emphasized that classical models require extensive feature engineering or hand-tuned pipelines, while neural networks can learn directly from the data and generalize more flexibly across tasks and domains.

Despite the promising results reported in many of these comparisons, a number of studies also highlighted the interpretability and reproducibility challenges inherent to ANN-based methods. As such, the integration of neural networks into the broader MEG analysis landscape is often framed not as a wholesale replacement of classical techniques, but as a complementary approach that may offer advantages under specific conditions. 

\section{Discussion}

\subsection{Added value of combining ANN with MEG}

ANNs are increasingly used alongside neuroimaging modalities such as fMRI, EEG, and MEG to enhance our understanding of brain function. fMRI is highly valued for its spatial resolution and ability to map cognitive states through blood flow changes (\cite{hall2014relationship}). However, its slower temporal resolution limits its capacity to capture rapid neural dynamics. In contrast, both MEG and EEG offer the temporal precision necessary for analyzing fast brain activities. However, MEG distinguishes itself with superior spatial resolution and reduced susceptibility to scalp and skull distortions (\cite{baillet2001electromagnetic, baillet2017magnetoencephalography, singh2014magnetoencephalography}).

Preprocessing steps, such as artifact removal and feature extraction, are crucial for ensuring that the high-dimensional MEG signals are effectively utilized by ANNs. These steps often form the foundation for downstream tasks by improving signal clarity and optimizing data input for architectures like CNNs, RNNs, and transformers (\cite{garg2017automatic, garg2017using, hyvarinen2016unsupervised}). In addition to preprocessing, ANNs are widely applied in MEG for classification tasks, including decoding (\cite{cichy2016comparison, abdellaoui2020deep}), BCIs (\cite{hramov2019kinesthetic, fan2022novel}), clinical diagnostics (\cite{guo2018stacked, guo2022transformer}), and event detection (\cite{zhao2022multi, hramov2018artificial}). By leveraging MEG's temporal precision, ANNs have been used to analyze neural synchronization patterns, facilitating insights into cognitive processes such as perception, attention, language processing, and decision-making. For example, in neural classification, MEG’s precise temporal information supports real-time detection of complex states (\cite{dash2020decoding, kim2019canet}); in BCI applications, it enhances user control through responsive spatiotemporal decoding (\cite{fan2022novel, hramov2019kinesthetic}); in clinical diagnostics, it enables more accurate seizure detection by addressing signal distortions; and in event detection (\cite{giovannetti2021deep, guo2018stacked}), it allows subtle oscillatory changes to be captured with higher fidelity (\cite{zhao2022multi, abdellaoui2021enhancing}).

Beyond these applications, MEG–ANN integration plays a critical role in modeling studies, where the goal is to explore how MEG signals align with representations learned by neural networks. These studies compare MEG data with internal ANN activations, providing insights into how cortical activity corresponds to hierarchical network processing (\cite{cichy2017dynamics, bankson2018temporal, seeliger2018convolutional}). MEG's high temporal and spatial resolution provides a unique advantage in tracking dynamic neural activity, making it particularly well-suited for investigating how the temporal dynamics of ANN activations align with neural processes in the human brain, especially in domains like vision and audition, where precise timing is critical (\cite{cichy2016comparison, caucheteux2022brains}).

\subsection{Current trends and dominant applications of ANNs in MEG} 

Taken together, the reviewed body of literature reveals a notable surge in publications and highlights the diverse applications of ANNs in MEG data analysis. This reflects not only a growing appreciation for data-driven approaches in neuroscience but also an increased interest in the unique contributions of ANNs to the field. Based on the papers we reviewed, it seems that the excitement about the potential of ANNs for advancing MEG research is distinct from the recent surge in using standard ML techniques (\cite{roy2019deep}). 

Our review shows that the power and versatility of deep learning is opening up frameworks for exploring MEG data that go beyond advanced statistical analyses. While many of the papers reviewed do indeed leverage the power of ANNs for classification tasks (\cite{cichy2016comparison, dash2020decoding, hramov2019kinesthetic}), many studies benefit from other strengths of ANNs. In particular, the reviewed literature shows that modeling studies that compare representations across ANNs and human MEG data seem to be gaining momentum (\cite{seeliger2018convolutional, donhauser2020two, wingfield2022similarities}). 

 Methodological developments for MEG data analytics are increasingly leveraging the strength and versatility of ANNs, including the introduction of novel architectures, enhanced data preprocessing techniques such as artifact detection and correction (\cite{hyvarinen2016unsupervised, croce2018deep}), innovative source reconstruction approaches (\cite{pantazis2021meg, sun2023deep}), and the development of foundation models (\cite{csaky2024foundational, wei2024nested, ortega2023brainlm}). These methodological innovations not only broaden the scope of MEG applications but also accelerate the pace of MEG research. That said, this review also highlights that there are still several critical limitations, including issues with interpretability and data scarcity, which will need to be addressed for the field to harness the potential of ANNs in MEG analysis fully. Further advancements —possibly including more robust foundation models for MEG that better handle limited datasets and enhance generalizability— are necessary to improve model transparency and effectively manage limited datasets to achieve more reliable and generalizable results. In the next section, we will explore in greater detail the bottlenecks and challenges that emerge from the corpus of studies included in this review.

\subsection{Current limitations and challenges}
The diversity and wide range of ways ANNs have been applied to MEG research reflect a correspondingly broad spectrum of challenges and limitations. To provide clarity, we categorize the predominant issues observed across the reviewed literature into a concise set of topics, offering strategic guidance where possible to enhance the robustness, efficacy, and reproducibility of future work in this rapidly evolving field. First, we address the general data scarcity problem prevalent in MEG-based neuroimaging studies. Next, we underscore the importance of model validation techniques and the need for accurate performance evaluation. We then discuss the critical role of hyperparameters in ANN-based MEG studies. Lastly, we examine the issue of reproducibility in this domain.

\subsubsection{Dealing with data scarcity}
It is recommended that data augmentation techniques be used when feasible to cope with scarce data. For instance, common approaches for time-series data like M/EEG include adding noise, applying transformations such as amplitude scaling or time warping, or generating synthetic trials using generative models. For event-based epochs, introducing different event timings for each trial window will increase the sample size. In the case of continuous data, it is possible to use overlapping segments to artificially increase dataset size. Depending on the type of study and data, various data augmentation techniques can be added to a study's preprocessing pipeline. For a systematic comparison and more details on available methods specifically evaluated on EEG, many of which are applicable to event-related MEG data, we refer the reader to \cite{rommel2022data}.

\subsubsection{Model validation techniques}
In many neural network applications, the vast amount of data combined with the typical training duration for ANNs makes cross-validation an impractical method for model validation. Instead, most of the time, the dataset is randomly split into three subsets to validate the model: training, validation, and test (also called evaluation set). The training set generally represents 50 to 80 percent of the initial dataset, and the remaining data is split in half to create the validation and test sets. The training set is employed to train the network with a set of hyperparameters and tested on the validation set. The test set is only used at the end when the architecture and hyper-parameters are fixed and once the desired performance has been reached on the validation set. It is solely used to evaluate the final model performance. However, when data is scarce, as it is often the case for neuroscience studies, it is recommended to use re-sampling methods such as bootstrapping or cross-validation to make use of the data as much as possible (\cite{riley2020calculating}). It will allow the train and validation sets to be changed several times to acquire a less noisy measure of the model's performance. In the studies included in this survey, when cross-validation is used, K-fold was the preferred technique, with the value of K ranging from 4 to 50. When the goal is to test generalization to new subjects, the leave P subjects out (LOPO) cross-validation or a stratified version of K-Fold is recommended.

\subsubsection{Model performance evaluation and baseline comparisons}
Some classification studies did not include baseline results against which one could have compared the reported ANN performances. This can be a relevant concern given that classical shallow ML approaches (e.g. logistic regression \cite{peng2002introduction}, random forest \cite{breiman2001random}, SVM \cite{boser1992training}, etc.) often involve simpler classifier implementations and potentially faster training times ---once appropriate features have been engineered---, compared to designing and training complex ANNs. However, the overall practical advantage depends heavily on the computational complexity of the necessary feature extraction pipeline, which can itself be intensive and time-consuming, potentially offsetting the gains in classifier simplicity. Additionally, small sample size can lead to inflated accuracies, that can substantially exceed the theoretical chance level of 50\% for two classes (eg. reaching 70\% purely by chance). Therefore, rigorous statistical evaluations, such as permutation tests are required to properly establish meaningful baseline performances, thus ensuring robust assessments of classifier significance \cite{combrisson2015exceeding}. Generally, one should always evaluate simple solutions before spending time and effort on more complex ones, which might also be more challenging to interpret. However, only a few articles among the modeling studies mentioned some kind of baseline performance comparison (\cite{dima2018spatial, bankson2018temporal, cichy2016comparison, cichy2017dynamics, van2022convolutional, caucheteux2022brains}). Ideally, modeling studies should report NC and/or evaluate the performance of their similarity analysis with an untrained model. Together, these measures would reinforce robust similarity scores' reliability and help identify and discard weaker results.

\subsubsection{Hyperparameter settings}
In recent years, we have witnessed the development of increasingly deep networks aimed, for a large part, at enhancing performance in image classification tasks. However, the studies reviewed here reveal a trend toward relatively shallower network architectures. Intriguingly, when depth is evaluated as a hyperparameter in these studies, it appears to have a lesser impact on performance compared to other parameters, such as layer width or filter size. When using ANNs, the training phase is a pivotal component, and an incorrect selection of the optimizer and parameters could result in a network that fails to learn. Unfortunately, 16 of the classification studies did not provide enough information about the hyperparameters used. However, most studies specify the loss function, optimizer, learning rates, batch size, and the number of epochs, including the early stop condition when applicable. It is essential to systematically detail a model's hyperparameters and training regimen (in any type of ANN study) to ensure the ability to reproduce results or to use the same model similarly with different data. In this context, a recommended practice is to also provide the source code alongside all the parameters and hyperparameters used to generate any given result of a study. Furthermore, given the dependence of optimal hyperparameters, including network depth, on the specific application, researchers designing networks for their own MEG studies may find the detailed information collated in tables \ref{tab:classification}, \ref{tab:classification2}, \ref{tab:modeling}, and \ref{tab:other} useful as a starting point or source of inspiration. Lastly, it's worth mentioning that many studies in our review employ batch sizes that are not powers of two. However, using batch sizes that are powers of two is recommended to optimize computation and reduce training times (\cite{kandel2020effect}).

\subsubsection{Variability in MEG data acquisition, preprocessing, and experimental protocols}
One of the main challenges in applying ANNs to MEG research is the variability in acquisition hardware, preprocessing steps, and experimental protocols across studies. These differences can affect data quality and consistency, which in turn influence model performance and the ability to compare results across studies.

At the acquisition level, studies used different sensor types (e.g., magnetometers, planar gradiometers, axial gradiometers) and MEG systems (e.g., Elekta Neuromag, CTF, BTi), with sampling frequencies ranging from 50 Hz to 2400 Hz. Preprocessing methods also varied: some studies used ICA (often via EEGLAB or FieldTrip) for artifact rejection, others applied wavelet denoising or relied on proprietary tools. In 39 studies, preprocessing steps were not clearly reported. Segment lengths and filtering parameters also differed, even in studies using similar tasks.

Experimental protocols varied in terms of task design (e.g., resting-state, oddball, language processing, or BCI tasks), number and timing of trials, baseline usage, and whether the data were epoched or treated as continuous. This variability makes it difficult to directly compare ANN architectures or performance metrics.

To address these challenges, future studies would benefit from more consistent preprocessing pipelines, standardized task protocols where feasible, and clearer reporting of methods. Efforts such as the brain imaging data structure extension for MEG (MEG-BIDS \cite{niso2018meg}) and open datasets with thorough documentation can support better reproducibility and enable more reliable benchmarking across studies. 

\subsubsection{The issue of reproducibility}
In the field of neuroscience, concerns about the reproducibility of results are prevalent. The substantial costs and privacy issues related to acquiring neuroimaging or behavioral data restrict the availability of open-access datasets, thereby challenging the reproducibility of research conducted with such data (\cite{poldrack2017scanning, button2013power}). In addition, this challenge is compounded by numerous studies that did not include enough information about the network architecture or the training parameters or did not provide truly open access to their code, from which this information could be extracted. While some studies state code is 'available upon request', this often proves ineffective in practice, with requests frequently going unanswered or being declined (\cite{gabelica2022many}), meaning truly accessible code is likely rarer than implied. This overall lack of transparency and accessibility further impedes the reproducibility of the results, making it difficult for researchers to validate and build on previous findings. The application of ML to neuroscience data faces challenges, mainly due to the relatively small/moderate sample sizes in neuroscience datasets. Besides, most algorithms require large amounts of data to achieve robust performance. Moreover, small datasets will make generalization harder to reach since having fewer data will further increase the impact of inter-subject variability. Solutions exist to tackle problems caused by small datasets, such as cross-validation, bootstrapping, and data augmentation. Still, more data will always be a preferable approach to achieve better results (\cite{dlbook}). However, the trend of small private datasets in the world of neuroscience is slowly changing with increasing awareness around the importance of open science practices, and public datasets are becoming more common. 

The utility of these public datasets for reproducible research and cross-study comparisons is greatly enhanced by standardized data formats and organizational structures, such as MEG-BIDS (\cite{niso2018meg}). Resources providing standardized, open-access MEG recordings suitable for ANN analysis include repositories and datasets such as OpenNeuro (\url{https://openneuro.org/}), the human connectome project (\url{https://www.humanconnectome.org/}), the Cambridge centre for ageing and neuroscience (Cam-CAN) dataset (\url{https://camcan-archive.mrc-cbu.cam.ac.uk/dataaccess/}), the open MEG archive (OMEGA) database (\url{https://www.mcgill.ca/bic/neuroinformatics/omega}), and the MNE sample datasets (\url{https://mne.tools/stable/overview/datasets_index.html}), offering valuable benchmarks for training, testing, and comparing models. 

\subsubsection{Interpretation and visualizations}
In BCI studies, where the primary concern is the algorithm's performance, the black-box nature of ANNs is often considered less problematic. However, in many other applications, including brain decoding studies, gaining insight into what is happening in the model's latent space is key to understanding how the information was extracted from the signal, as well as why and how decoding is possible. In domains like image classification or object detection, there are various tools (Grad-CAM \cite{selvaraju2020grad}, tSNE \cite{maaten2008visualizing}, UMAP \cite{mcinnes2018umap}) that allow the user to have a representation of what the network 'sees' and what properties of the data are important in its decision-making process. 

A large proportion (52 out of 70) of the classification studies, where interpretation would be the most interesting, did not use these tools to gain insights into the representations learned and used by the model. A similar pattern of limited application was observed in the 'Other' category studies reviewed. While interpretability might be considered less critical for some preprocessing or source localization tasks compared to decoding, it remains vital for evaluating novel methodological approaches. Indeed, roughly half of the 'Methods' studies (9 out of 16) included visualization techniques such as filter or activation map analysis. Interpretability tools were also used in a majority of the 'Preprocessing' studies (5 out of 8), often employing methods like Grad-CAM \cite{selvaraju2017grad}. However, interpretability techniques were notably absent in all reviewed ANN-based source localization studies. This suggests that developing and applying robust interpretability methods beyond classification and basic modeling tasks remains an important area for future work in MEG research using ANNs.

\subsubsection{Note of caution on inconsistent nomenclature}
Consistent terminology is important for effective communication across all scientific disciplines. However, it becomes particularly crucial in emerging fields where the community still defines methodologies and core concepts. Adopting a uniform nomenclature is essential in a rapidly evolving research domain, such as the intersection of ANN and neuroimaging. Our extensive literature review reveals that terminology within classification studies is fairly consistent. However, we noted that the nomenclature in other categories, particularly within modeling studies, tends to show less consistency. To give a few examples, some research papers include RSA as a component of the MVPA analysis, while others do not. Additionally, some studies use neural predictability but do not explicitly mention it in their papers. In the modeling studies reviewed, the analysis pipelines exhibit considerable variety and flexibility, leading to notable diversity across the studies. For researchers looking to adopt these methodologies, we recommend a thorough examination of multiple studies. This approach will provide a comprehensive understanding of the various pipelines and help identify the methods most aligned with specific research objectives. Additionally, it is crucial to strive for consistency in terminology with the existing literature to ensure clarity and help the community move towards established procedures and nomenclature. For those interested in RSA in particular, insightful comments and critiques of the method can be found in \cite{dujmovic2022some}.

\subsubsection{Hardware and practical application constraints}
Another consideration involves the practical constraints of MEG for certain applications, notably BCIs. While EEG is often considered more practical for widespread BCI use due to lower cost and portability, MEG's arguably superior signal quality and spatial resolution offer potential advantages for specific BCI paradigms requiring high fidelity decoding, as explored in several studies reviewed here (\cite{dash2020decoding, yeom2020lstm, zubarev2019adaptive}). However, traditional SQUID-based MEG systems face significant hurdles for practical BCI deployment, including high cost, the need for magnetically shielded rooms, and strict subject immobility requirements. Although ANNs can contribute by improving the robustness of decoding complex MEG signals and potentially adapting to noise or variability (\ref{sec:classif}), they do not resolve these fundamental infrastructural and hardware limitations. Promisingly, emerging sensor technologies like optically pumped magnetometers (OPMs) may mitigate these physical constraints. OPMs operate without cryogenics, are wearable, and allow for subject movement, potentially enabling more flexible, cost-effective, and practical MEG-based BCI systems in the future (\cite{fedosov2024low}).

\subsubsection{Challenges in ANN-based source reconstruction}
Applying ANNs to the MEG inverse problem, i.e., source reconstruction, presents both opportunities and significant challenges, as highlighted by the studies reviewed in section \ref{sec:other}. Source reconstruction is inherently ill-posed and highly sensitive to factors like sensor noise, assumptions about source activity, and, crucially, the accuracy of the head model used to compute the forward solution (\cite{baillet2001electromagnetic, hamalainen1993magnetoencephalography}). While ANNs offer potential advantages, such as learning complex non-linear source-sensor mappings directly from data, their effectiveness, particularly when trained on simulated data as is common practice (\cite{pantazis2021meg, sun2023deep}), is fundamentally constrained by the fidelity of these simulations. 

Indeed, if an ANN is trained using data generated with an inaccurate forward model (due to simplified head geometry, incorrect conductivity values, or sensor misalignment), its ability to generalize and accurately localize sources in real-world MEG data will be severely compromised. The network may perform well on simulated test data derived from the same flawed model but fail when applied to actual measurements. Therefore, the development of effective ANN-based source reconstruction methods heavily relies on the generation of high-quality training data using realistic simulations and, most importantly, accurate subject-specific forward models. Furthermore, even with accurate models, challenges remain in ensuring that ANNs do not simply learn biases present in the simulation protocols and in interpreting the learned source representations. It is therefore important to keep in mind that although ANNs may provide a powerful data-driven alternative to traditional inverse methods, they do not circumvent the fundamental physical constraints and modeling requirements inherent to MEG source localization. 

\section{Conclusion}

Machine learning, notably the use of ANNs, has become a cornerstone in contemporary research, reshaping both data analysis and modeling across diverse scientific fields. Although ANNs are relatively novel in the realm of neuroimaging, the collection of studies in this review demonstrates a rapidly growing interest in employing ANNs for both the analysis and modeling of MEG data. This trend is part of a broader movement towards integrating advanced data-driven and computational methods to deepen our understanding of neural mechanisms. The large body of work we reviewed indicates that ANNs are increasingly recognized as powerful tools in neuromagnetic imaging due to their robust performance, versatility, and ability to handle complex datasets. A comparison between our figure \ref{fig:years} and figure 10b in \cite{roy2019deep} suggests a trajectory similar to that observed in the field of EEG, where ANNs have already shown extensive utility. This parallel points to a likely expansion in the number of studies employing ANNs for MEG data, further diversifying the applications and methodologies within the field. 

One of the most remarkable observations is the diverse range of applications for which ANNs are now being employed with MEG data. While traditional ML tools like SVM or random forest classifiers are well-established, particularly for brain decoding and classification tasks based on extracted features (\cite{mlneuro}), ANNs facilitate a distinct spectrum of applications, as demonstrated throughout this review. Notably, their capacity for representation learning and building end-to-end systems has led to their use not only in classification but also in addressing the source estimation problem (\cite{pantazis2021meg, sun2023deep}), providing novel tools for data cleaning and artifact rejection (\cite{treacher2021megnet, hamdan2023reducing}), and, significantly, building and evaluating complex information processing models of the human brain (\cite{cichy2017dynamics, wingfield2022similarities, yang2020artificial}). 

Despite the remarkable advances in AI tools and the new opportunities that ANNs present for MEG data analysis and modeling, it is critical to avoid the allure of AI hype and the rush to employ complex AI algorithms solely for their novelty. Many research questions may still be best addressed using standard approaches that are well-established within the field and which often come with greater result interpretability. Moreover, the extensive range of parameters and hyperparameters involved in ANNs, coupled with the expertise required for their proper implementation, can sometimes introduce errors or yield misleading outcomes. 

Looking ahead, the future of this emerging research appears ripe with promising developments. Two particularly exciting prospects stand out. First, advances in AI interpretability and visualization tools are poised to significantly enhance the application of ANNs in MEG research, making these complex models more accessible and their findings more actionable. Second, the burgeoning adoption of AI foundation models across various disciplines, including neuroscience, is set to play a transformative role in both basic and clinical MEG applications. The scientific community's growing enthusiasm for developing foundation models for brain imaging modalities (\cite{ortega2023brainlm, wang2023large}) and time series data (\cite{garza2023timegpt}) suggests that neuroscience will not be far behind (\cite{csaky2024foundational, wei2024nested}). Several studies have successfully applied ANNs to multimodal neuroimaging data, including MEG-EEG integration. For example, frameworks have been developed to process and fuse MEG and EEG signals for enhanced decoding performance and artifact removal using deep neural architectures (\cite{o2024localized, zubarev2022mneflow, dehgan2025meegnet}). These approaches highlight the potential of ANNs to handle the complementary strengths of different modalities and suggest promising avenues for future multimodal brain decoding pipelines. Integrating such modalities could revolutionize our understanding of the brain and its disorders.

\newpage

\section*{Acknowledgements}

\subsection*{Conflict of interest}
The authors declare that they have no conflicts of interest relevant to this work.

\subsection*{Funding}
I.R. received support from the Canada CIFAR AI chair program and the Canada excellence research chairs program. K.J. was supported by Canada research chairs program funding (950-232368). V.H. was supported through the ianadian institutes of health tesearch project grant (166197). H.A. was funded through the faculty of medicine at the university of Montreal's merit scholarship. A.D. was funded through computational neuroscience and cognitive neuroimaging lab (CoCoLab) and Montreal institute of learning algorithms (MILA) scholarships.


\end{document}